\documentclass[12pt]{article}
\usepackage[dvips]{epsfig}

\font\mybb=msbm10 at 12pt

\font\mybbsub=msbm10 at 8pt
\font\mybbsmall=msbm10 at 10pt

\def\bb#1{\hbox{\mybb#1}}

\def\bbsub#1{\hbox{\mybbsub#1}}
\def\bbsmall#1{\hbox{\mybbsmall#1}}

\def\ZZ {\bb{Z}}
\def\ZZsub {\bbsub{Z}}
\def\ZZsmall {\bbsmall{Z}}

\def\RR {\bb{R}}

\def\A {\mbox{\scriptsize A}}

\newcommand{\beq}{\begin{equation}}
\newcommand{\eeq}{\end{equation}}
\newcommand{\beqa}{\begin{eqnarray}}
\newcommand{\eeqa}{\end{eqnarray}}
\newcommand{\bcen}{\begin{center}}
\newcommand{\ecen}{\end{center}}
\newcommand{\btab}[1]{\begin{tabular}{#1}}
\newcommand{\etab}{\end{tabular}}

\newcommand{\non}{\nonumber}

\newcommand{\common}{\mbox{\scriptsize \boldmath $p$}^2
                    +\mbox{\scriptsize \boldmath $pq$}
                    +\mbox{\scriptsize \boldmath $q$}^2}
\newcommand{\p}{\mbox{\boldmath $p$}}
\newcommand{\q}{\mbox{\boldmath $q$}}
\newcommand{\psub}{\mbox{\scriptsize \boldmath $p$}}
\newcommand{\qsub}{\mbox{\scriptsize \boldmath $q$}}
\newcommand{\ptiny}{\mbox{\tiny \boldmath $p$}}
\newcommand{\qtiny}{\mbox{\tiny \boldmath $q$}}

\begin{document}
\begin{titlepage}

\setcounter{page}{0}
\begin{flushright}
KEK Preprint 2001-24 \\
hep-th/0105174
\end{flushright}

\vskip 5mm
\begin{center}
{\Large\bf Wound D-branes in Gepner Models\\}
\vskip 10mm

{\large
Shun'ya Mizoguchi\footnote{\tt mizoguch@post.kek.jp}
and Taro Tani\footnote{\tt tanitaro@post.kek.jp}
} \\
\vskip 5mm
{\it Institute of Particle and Nuclear Studies \\
High Energy Accelerator Research Organization (KEK) \\
Oho 1-1, Tsukuba, Ibaraki 305-0801, Japan} \\
\end{center}
\vskip 10mm
\centerline{{\bf{Abstract}}}
\vskip 3mm

We propose a new prescription of how to represent D-branes in Gepner
models in more general homology classes than those in the previous
constructions. The central role is played by a certain projection acting
on the Recknagel-Schomerus boundary states. Consequently, the boundary
states are in most cases no longer a sum of products of $N=2$ Ishibashi
states, but nevertheless preserve spacetime supersymmetry and satisfy 
the Cardy condition. We demonstrate these in the $(k=1)^3$ Gepner model 
in detail, and construct boundary states for D-branes wound around
arbitrary rigid 1-cycles on the corresponding 2-torus. We also emphasize
the necessity of some angle-dependent transformations in identifying a
proper free-field  realization for each brane tilted at an angle. In
particular, this is essential for the Witten index to give the correct
intersection numbers between the different D-branes. 

\end{titlepage}
\baselineskip=18pt
\setcounter{footnote}{0}
\setcounter{equation}{0}
\section{Introduction}
~~~
One of the important issues in the study of D-branes is their stability 
in Calabi-Yau compactifications. 
(See \cite{Douglas,Douglas2} for a quick  overview on the subject.)
In general, a spectrum of BPS states in a weakly coupled theory is not
necessarily the same as the one at strong couplings, as we know from
many field-theory examples.  Relations between D-branes among  
different points in Calabi-Yau moduli space were first studied in the
quintic case \cite{D-branes_on_the_quintic} by utilizing the knowledge 
of monodromies of the periods \cite{CDGP} and the boundary states in
Gepner models \cite{RS}. Related developments may be found in
\cite{DG}-\cite{DFR}.

Gepner models \cite{Gepner} are a useful way to describe a 
compactification on a specific Calabi-Yau manifold which (typically)
has a small volume. Therefore, the boundary-state representation of
D-branes in Gepner models, pioneered by Recknagel and Schomerus
\cite{RS}, provides us a tractable framework for analyzing D-branes in
the stringy (strong sigma-model coupling) regime.
(Earlier seminal and other related works are
\cite{Cardy}-\cite{BD}.)

Although very useful, their proposed boundary states for D-branes at 
the Gepner point, called {\it rational} boundary states, do not exhaust 
all possible supersymmetric D-branes \cite{BBS} in the Calabi-Yau space.
For example, in the ($k$=1)${}^3$ Gepner model, which describes  a
compactification upon a certain flat 2-torus, it is known
\cite{GS} that the Recknagel-Schomerus A-boundary states can represent
{\it only}  D1-branes
\footnote{For definiteness, we assume here that there are no spacelike
noncompact Neumann directions.}
wrapped around either of the shortest nontrivial
1-cycles. 
Since it is straightforward (as we see in section 3) to represent
D1-branes wound around arbitrary rigid 1-cycles in terms of the toroidal
CFT, our present  technology of constructing D-brane boundary states in 
Gepner models is clearly unsatisfactory. 
\footnote{Also, in the quintic case, pure D0-branes have no 
corresponding rational boundary states at the Gepner point, but this does
not necessarily mean their nonexistence \cite{D-branes_on_the_quintic}.
(See \cite{recent} for more recent discussions.)
Another example is the D-branes wrapping the cycles induced by the
resolution of the singularity \cite{BS}.}

In this paper,
we give an answer to the basic question of how these infinitely many 
D1-branes wound around various 1-cycles are described as boundary 
states in terms of the Gepner-model language. Our strategy is to 
first calculate the open-string partition functions between parallel 
and intersecting D1-branes at arbitrary angles on a general 2-torus by 
means of the toroidal CFT approach, and then construct boundary states in
the ($k$=1)$^3$ Gepner model so that the partition functions between them
coincide with the `geometric' partition functions for the 
corresponding 2-torus.

Although we discuss in detail only this simplest example of Gepner
models, it already presents the essential features and immediately
indicates how we may generalize the Recknagel-Schomerus states in more
complicated Gepner models which correspond to more general Calabi-Yau
compactifications. Some of the notable features of our construction 
include:
\begin{itemize}
\item
A projection operation acting on a Recknagel-Schomerus state, which
leaves only a subsector satisfying a certain $U(1)$-charge condition;
this  condition is linear and {\it orthogonal} to the total $U(1)$-charge
condition relevant to supersymmetry. Consequently the resulting new
boundary states are also supersymmetric.
\item
A treatment of partition functions in terms of $\beta$-orbits; 
this allows us to easily (or rather automatically) incorporate 
the Cardy condition (integrality of the coefficients of the open-channel
partition function), as well as exhibits manifest supersymmetry.
\item
A suitable identification of a proper free-field realization for each
brane tilted at an angle; to distinguish one tilted D1-brane from
another,  it turns out that one must identify a set of free fields
realizing $N=2$ minimal models for each brane through some {\it
angle-dependent transformations}. 
\end{itemize}
Owing to the projection, the boundary states are in general no longer a
sum of products of $N=2$ Ishibashi states.

At this point, we would like to clarify the difference between our 
prescription and the work \cite{FSW}, where the construction in
\cite{RS} was extended by twisting the `simple currents' 
(the spectral-flow and the `fermion-aligning' operators) as well as taking
into account their `fixed point', and the expressions of 
\cite{RS,BS} were 
recovered as special cases. 
As mentioned there, the former amounts to shifting the modulus of the
total $U(1)$ current; such a shift also appears in this paper as a part of
the angle-dependent transformation. The most significant difference is
that our prescription includes a projection with respect to another
$U(1)$ which is {\it orthogonal} to the total $U(1)$ current; here the
order of the cyclic group for the projection may be arbitrarily large 
depending on the length of the cycle.
Another new aspect of our prescription is the angle-dependent rotation
(which is also orthogonal to the total $U(1)$) in identifying the proper
free-field realization for each tilted D1-brane. We show that this is
indispensable for the Witten index to give the correct intersection
numbers between different D1-branes.

This paper is organized as follows.
In section 2, we give a brief summary of the known construction of
boundary states in general Gepner models. In section 3, we calculate 
the open-string partition functions in the toroidal CFT formulation.
In section 4, we construct boundary states in the ($k$=1)$^3$ Gepner model
which reproduce the geometric partition functions and the intersection 
numbers. In section 5, we outline the construction of boundary states 
in the ($k$=2)$^2$ model.
The last section summarizes our results and also includes a
discussion on the application to general Gepner models.

\setcounter{equation}{0}
\section{Boundary states in Gepner models}
~~~
We begin by briefly reviewing how D-branes are 
described in Gepner models \cite{RS}.  
Consider a Gepner model defined by $r$ tensor
product of $N=2$ minimal models with level 
$k_j$ $(j=1,\ldots,r)$, describing a compactification of 
type II string theory to $(d+2)$-dimensional spacetime. 
Although we will mainly consider the case
for $d=6$, the case for $d=2$ can be treated in parallel
as well. We assume $d=2$ or 6 in the following.

Recknagel and Schomerus assumed that the 
A-boundary states in this model are given by \cite{RS} 
\begin{eqnarray}
|\alpha\!>_{\A}&\equiv&|S_0;(L_j,M_j,S_j)\!\!>_{\A}\nonumber\\
&=&\frac1{\kappa_\alpha^A}
\sum_{\lambda,\mu}^\beta
B^{\lambda,\mu}_\alpha|\lambda,\mu\!>\!\!>_{\A},
\label{RSboundarystates}
\end{eqnarray}
\begin{eqnarray}
B^{\lambda,\mu}_\alpha&=&
(-1)^{\frac{s_0^2}2}
e^{-\pi i\frac{ds_0S_0}4} \prod_{j=1}^r\left(
\frac{\sin\pi\frac{(l_j+1)(L_j+1)}{k_j+2}}
{\sin^{1/2}\pi\frac{l_j+1}{k_j+2}}
e^{\pi i\left(
\frac{m_jM_j}{k_j+2}-\frac{s_jS_j}2
\right)}
\right).\label{B^lambdamu_alpha}
\end{eqnarray}
$|\lambda,\mu>\!\!>_{\A}$ is a product of $r$ $N=2$ A-type Ishibashi 
states 
\begin{eqnarray}
|\lambda,\mu\!>\!\!>_{\A}&=&|s_0\!>\!\!>
\prod_{j=1}^r|l_j,m_j,s_j\!>\!\!>_{\A}
\end{eqnarray}
with the collective labels 
\begin{eqnarray}
\lambda=\{l_1,\ldots,l_r\}, ~~~~~~
\mu=\{s_0~;m_1,\ldots,m_r;s_1,\ldots,s_r\},
\end{eqnarray}
where
\begin{eqnarray}
l_j=0,\ldots,k, ~~~m_j\in \ZZ_{2(k_j+2)},~~~ s_j\in \ZZ_4~~~ 
(j=1,\ldots,r) \label{range}
\end{eqnarray}
are the labels of the irreducible representations of the $N=2$ 
superconformal algebra. \mbox{$s_0\in \ZZ_4$} labels the irreducible
representations of the $SO(d)$ current algebra. 
The normalizations of the Ishibashi states are  
\begin{eqnarray}
&&{}_{\A}\!<\!\!<\tilde{l},\tilde{m},\tilde{s}~
|\tilde{q}^{L^{N=2}_0-\frac{c^{N=2}}{24}}
|l,m,s\!>\!\!>_{\A}
\nonumber\\&&~~~~~~~~~~~~~~~~~~~~~~~~~~~~
~~=~\left(
\delta_{\tilde{l},l}
\delta^{(2(k+2))}_{\tilde{m},m}
\delta^{(4)}_{\tilde{s},s}
+\delta_{\tilde{l},k-l}
\delta^{(2(k+2))}_{\tilde{m},m+k+2}
\delta^{(4)}_{\tilde{s},s+2}
\right)
\chi^{l,s}_m(\tilde\tau,0),
\label{product}\\
&&<\!\!<\tilde{s}_0|
\tilde{q}^{L_0^{SO(d)}-\frac{d}{48}}|
s_0>\!\!>
~=~\delta_{\tilde{s}_0,s_0}^{(4)}
\chi^{SO(d)}_{s_0}(\tilde\tau),
\end{eqnarray}
where 
$\chi^{l,s}_m(\tilde\tau,z)$ and $\chi^{SO(d)}_{s_0}(\tilde\tau)$
$(\tilde{q}=e^{2\pi i\tilde\tau})$ 
are the $N=2$ minimal and the $SO(d)$ characters, 
respectively. 
$\chi^{l,s}_m(\tilde\tau,z)=0$ if $l+m+s\neq 0$ mod 2.
$\delta^{(N)}_{\tilde{m},m}$ is the delta function on $\ZZ_N$.

The `$CPT$-conjugate'  boundary state is defined by
\begin{eqnarray}
{}_{\A}<\Theta\alpha|&\equiv&
\frac1{\kappa_\alpha^A}
\sum_{\lambda,\mu}^\beta
B^{\lambda,\mu}_\alpha{}_{\A}\!\!<\!\!<\!\lambda,-\mu|.
\end{eqnarray}
This operation is almost equivalent to taking the hermitian conjugate,
with the exception that the factor $(-1)^{s_0^2/2}$ remains
unchanged; it is so arranged that the partition function with 
itself becomes an alternating summation (and hence supersymmetric).
The partition function between the two boundary states 
$\alpha$ and $\tilde{\alpha}$ is calculated as follows \cite{RS}:
\begin{eqnarray}
Z^A_{\tilde{\alpha}\alpha}(\tau)
&\equiv&{}_{\A}\!<\Theta\tilde{\alpha}|\tilde{q}^{L_0-\frac
c{24}}|\alpha >_{\A}
\nonumber\\
&=&\frac{2^r}{\kappa_{\tilde\alpha}^A\kappa_\alpha^A}
\sum_{\lambda,\mu}
^\beta
B^{\lambda,-\mu}_{\tilde\alpha}
B^{\lambda,\mu}_\alpha
\chi^{\lambda,\mu}(\tilde\tau)
\nonumber\\
&=&\frac{2^r}{\kappa_{\tilde\alpha}^A\kappa_\alpha^A}
\sum_{\lambda,\mu}^\beta
\sum_{\lambda',\mu'}
B^{\lambda,-\mu}_{\tilde\alpha}
B^{\lambda,\mu}_\alpha
S^{SO(d)}_{s_0,s'_0}
\left(\prod_{j=1}^r S^{(k_j)}_{(l_j,m_j,s_j),(l'_j,m'_j,s'_j)}
\right)
\chi^{\lambda',\mu'}(\tau)
\nonumber\\
&=&\sum_{\lambda',\mu'}\sum_{\nu_0=0}^{K-1}\sum_{\nu_1\ldots,\nu_r=0,1}
(-1)^{\nu_0}\delta^{(4)}_{s_0',
								2+\tilde{S}_0-S_0-\nu_0-2\sum_{j=1}^r\nu_j}\nonumber\\
&&\cdot\prod_{j=1}^r\left(
N_{L_j\tilde{L}_j}^{l_j'}
\delta^{(2(k_j+2))}_{m_j',\tilde{M}_j-M_j-\nu_0}
\delta^{(4)}_{s_j',\tilde{S}_j-S_j-\nu_0+2\nu_j}
\right)
\chi^{\lambda',\mu'}(\tau),\label{Z_alphaalphatilde}
\end{eqnarray}
where $\tilde\tau\equiv -1/\tau$,
$K\equiv \mbox{lcm}(4,2(k_j+2))$ and
\begin{eqnarray}
{}_{\A}\!\!<\!\!<\!\lambda,\mu|~\tilde{q}^{L_0-\frac c{24}}
|\lambda,\mu>\!\!>_{\A}
&=&\chi^{\lambda,\mu}(\tilde\tau)\nonumber\\
&\equiv&
\chi^{SO(d)}_{s_0}(\tilde\tau)
\prod_{j=1}^r\chi^{l_j,s_j}_{m_j}(\tilde\tau,0).
\end{eqnarray}
$\chi^{\lambda',\mu'}(\tau)$ is also given by an obvious formula.
$S^{SO(d)}_{s_0,s'_0}$,
$S^{(k)}_{(l,m,s),(l',m',s')}$
defined by 
\footnote{
There is a typo in the modular transformation 
in \cite{Gepner}, and that has also been transmitted to many places in
the literature; the correct prefactor of $S^{(k)}_{(l,m,s),(l',m',s')}$ 
is $\frac1{2(k+2)}$ \cite{RY} and not $\frac1{\sqrt{2}(k+2)}$. }
\begin{eqnarray}
S^{SO(d)}_{s_0,s'_0}&=&\frac12e^{-\frac14\pi ids_0s'_0},
\nonumber\\
S^{(k)}_{(l,m,s),(l',m',s')}&=&
\frac1{2(k+2)}\sin\frac{\pi(l+1)(l'+1)}{k+2}
e^{\pi i(\frac{mm'}{k+2}-\frac{ss'}2)}
\end{eqnarray}
are the modular transformation matrices
\begin{eqnarray}
\chi^{SO(d)}_{s_0}\left(-\frac1\tau\right)
&=&\sum_{s'_0\in \ZZsub_4}S^{SO(d)}_{s_0,s'_0}\chi^{SO(d)}_{s'_0}(\tau),
\nonumber\\
\chi^{l,s}_m\left(
-\frac1\tau,\frac z\tau\right)
&=&\sum_{l'=0}^k\sum_{s'\in \ZZsub_4}
\sum_{m'\in \ZZsub_{2(k+2)}}\!\!\!\!\!
S^{(k)}_{(l,m,s),(l',m',s')}e^{\frac k{k+2}\pi i\frac{z^2}\tau}
\chi^{l,s}_m(\tau,z).
\label{modular}
\end{eqnarray}
$N_{L_j\tilde{L}_j}^{l_j'}$is the $SU(2)$ fusion coefficients. 
Finally, we have
chosen the normalization constant 
$\kappa^A_\alpha$ as
\begin{eqnarray}
\kappa^A_\alpha&=&
\left(
\frac{2^{r+1}\prod_{j=1}^r(k_j+2)}{K}
\right)^{1/2}
\end{eqnarray}
in the last expression.

An important observation is that the alternating summation over $\nu_0$
of the product of characters (\ref{Z_alphaalphatilde}) is precisely the
spectral-flow ($\beta$-) orbit, which was originally introduced
\cite{Gepner} in the construction of modular invariant partition
functions for closed superstring compactifications. The partition
function 
$Z^A_{\tilde{\alpha}\alpha}(\tau)$ 
vanishes if the `initial condition' of the flow 
\begin{eqnarray}
(\tilde{S}_0-S_0+2; \tilde{M}_j-M_j, \tilde{S}_j-S_j) 
\label{allnu_j=0}
\end{eqnarray}
satisfies the $\beta$-condition
\begin{eqnarray}
-\frac d8(\tilde{S}_0-S_0)
+\frac 12 \sum_{j=1}^r \left(
\frac{\tilde{M}_j-M_j}{k_j+2}
-\frac{\tilde{S}_j-S_j}{2}
\right)
&\in& \ZZ
\label{beta-condition}
\end{eqnarray}
for $d=2$ or 6.
If (\ref{allnu_j=0}) satisfies
(\ref{beta-condition}), then so do all the other orbits in 
$Z^A_{\tilde{\alpha}\alpha}(\tau)$ 
automatically, and hence it is 
manifestly supersymmetric. 
In particular, if $\alpha=\tilde{\alpha}$, then 
\begin{eqnarray}
Z^A_{\alpha\alpha}(\tau)
&=&\sum_{\lambda',\mu'}\sum_{\nu_0=0}^{K-1}\sum_{\nu_1\ldots,\nu_r=0,1}
(-1)^{\nu_0}\delta^{(4)}_{s_0',
								2-\nu_0-2\sum_{j=1}^r\nu_j}\nonumber\\
&&\cdot\prod_{j=1}^r\left(
N_{L_jL_j}^{l_j'}
\delta^{(2(k_j+2))}_{m_j',-\nu_0}
\delta^{(4)}_{s_j',-\nu_0+2\nu_j}
\right)
\chi^{\lambda',\mu'}(\tau). \label{Z_alphaalpha}
\end{eqnarray}
The open-channel partition function between identical boundary 
states is thus given by the sum of $\beta$-orbits with labels  
\begin{eqnarray}
(2-2\sum_j\nu_j; 0, 2\nu_j) .
\end{eqnarray}

Since (\ref{Z_alphaalphatilde}) depends on 
$(S_0; M_j, S_j)$,
$(\tilde{S}_0; \tilde{M}_j, \tilde{S}_j)$ only through 
their differences,
the partition
function for a  boundary state with itself always reduces to
(\ref{Z_alphaalpha}),  irrespective of the specific values of 
$(S_0; M_j,S_j)$.  This means that the boundary states of the type
(\ref{RSboundarystates}) can represent  only finitely many D-branes
in  different geometric configurations.

\setcounter{equation}{0}
\section{Partition functions of D-branes on general tori}
~~~
In this section, we derive the partition functions 
between arbitrary pairs of D-branes on a general 2-torus.
There are two types of partition functions
depending on the relative positions of D-branes: 
parallel or tilted at an angle.
We construct boundary states in the $(k=1)^3$ Gepner model
in section 4 so that they reproduce the geometric 
partition functions given in this section
for the corresponding torus ($SU(3)$ torus).
\subsection{Parallel branes}
~~~
In this subsection, we derive
the partition functions for parallel D$p$-branes
on a 2-torus ($\times$ eight-dimensional Minkowski space)
and find the corresponding geometric boundary states.
We assume that one of the $p$-dimensional worldvolume winds
around a 1-cycle labeled by the relatively prime
winding numbers $(\p,\q)$ (Figure 1).
\vspace{0.3cm}\\
{\it The open-string channel} \vspace{0.2cm} 

The partition function is defined in the open-string channel
by a one-loop amplitude
\beq
Z^{\mbox{\scriptsize{tot}}}= 
           [\mbox{log}(\mbox{det}
            H^{(\mbox{\scriptsize{o}})})]^{-1/2}
         = \int_0^{\infty}\frac{dt}{2t}\mbox{Tr}\,
           e^{-2\pi tH^{(\mbox{\tiny{o}})}}
\label{eq:openpf},
\eeq
where $H^{(\mbox{\scriptsize{o}})}$ is the open-string 
hamiltonian,\, $\mbox{Tr}$ is the sum 
over the degrees of freedom of the open string 
ending on the two parallel branes.
In the following, we assume that the two branes
are on top of each other.

The calculation of the trace is straightforward
except for the summation of the momenta and winding modes. 
We first give the result:
\beqa
  Z^{\mbox{\scriptsize{tot}}}
    _{\theta=\theta_{\ptiny,\qtiny}}
   &=& 2V_p \int_0^{\infty}\frac{dt}{2t}
     (8\pi^2t \alpha')^{-\frac{p}{2}}
      \frac{1}{\eta^6 (\tau)}\non \\
 & &   \cdot \frac{1}{\eta^2(\tau)}
              Z^{\mbox{\scriptsize{open}}}
               _{\theta=\theta_{\ptiny,\qtiny},0}(\tau)
              \frac{1}{2}\frac{\vartheta_3^4(\tau,0)
                              -\vartheta_2^4(\tau,0)
                              -\vartheta_4^4(\tau,0)}
                              {\eta^{4}(\tau)}
\label{eq:openpfpq},
\eeqa 
with
\beq
Z^{\mbox{\scriptsize{open}}}
    _{\theta=\theta_{\ptiny,\qtiny},0}(\tau)
         = \sum_{m,n \in \ZZsub}q^{\frac{1}
           {G_{11}\psub^2+2 G_{12}\psub\qsub
           +G_{22}\qsub^2}
           (m^2-2 B_{12} mn + (G+B_{12}^2)n^2)}
\label{eq:momwindo}
\eeq
($\tau=it$). $2V_p$ and $(8\pi^2t\alpha')^{-p/2}$ come from
the integral of the coordinates and the momenta of 
the noncompact directions, 
while the modular forms are the oscillator sum
for all the transverse directions with the GSO projection.

Let us sketch the derivation of the formula (\ref{eq:momwindo}).
Let $G_{\mu \nu}$ and $B_{\mu \nu}$ be
the metric and the antisymmetric-tensor-field background
of the torus.
We normalize the metric so that
the radius of the $X^1$ direction is 
$\sqrt{G_{11}}\sqrt{\alpha'}$ 
and the volume of the torus is
$\sqrt{G}(2\pi \sqrt{\alpha'})^2$
($G$ is the determinant of the metric).
For example, the backgrounds corresponding to
the $SU(3)$ and $SU(2)^2$ tori in this unit are
\beqa
&  & SU(3)\,\mbox{torus}\,\,:\quad
        G_{11}=G_{22}=1, \quad G_{12}=B_{12}=1/2 , \non \\
&  & SU(2)^2\,\mbox{torus}:\quad
        G_{11}=G_{22}=1, \quad G_{12}=B_{12}=0 .  \non 
\eeqa
At the boundaries ($\sigma=0,\pi$),
the open string satisfies
\beq
\delta X^{\mu} (G_{\mu \nu}\partial_{\sigma}X^{\nu}
+B_{\mu \nu}\partial_{\tau}X^{\nu})=0,
\eeq
or equivalently
\beq
  \partial_{\tau} \widehat{X}^a_{\theta=0} \,
  \partial_{\sigma} \widehat{X}^a_{\theta=0}=0.
\label{eq:torusbco}
\eeq
Here we introduced the reference local Lorentz coordinates 
\beq
\widehat{X}^a_{\theta=0} 
= e^a_{\, \, \mu} X^{\mu} \quad (a=1,2),
\eeq
where the zweibein $e^a_{\, \, \mu}$ 
(and its inverse $e^{\mu}_{\, \, a}$) 
is chosen so that the directions of 
$\widehat{X}^1_{\theta=0}$ and $X^1$ coincide, {\it i.e.},
one of the sides of the torus lies on the 
$\widehat{X}^1_{\theta=0}$ direction:
\beq
e^a_{\,\, \mu}=\frac{1}{\sqrt{G_{11}}}
                 \left(
                 \begin{array}{cc}
                 G_{11}  &  G_{12} \\
                 0       &   \sqrt{G}
                 \end{array}
                 \right), \quad  
e^{\mu}_{\,\, a}=\frac{1}{\sqrt{G G_{11}}}
                 \left(
                 \begin{array}{cc}
                 \sqrt{G}    &  - G_{12} \\
                 0           &    G_{11}
                 \end{array}
                 \right).
\eeq
The boundary condition (\ref{eq:torusbco})
is decomposed for the $(\p,\q)$ brane
in terms of the rotated coordinates as
\beq
 \partial_{\sigma}\widehat{X}^1_{\theta=\theta_{\ptiny,\qtiny}} 
                            =0\, , \quad
 \partial_{\tau}\widehat{X}^2_{\theta=\theta_{\ptiny,\qtiny}}
                            =0,
\label{eq:openbc}
\eeq
where 
\beq
    \left( \begin{array}{c}
       \widehat{X}^1_{\theta=\theta_{\ptiny,\qtiny}}  \\
       \widehat{X}^2_{\theta=\theta_{\ptiny,\qtiny}} 
            \end{array}
     \right)
            = R(\theta_{\psub,\qsub})
               \left( \begin{array}{c}
                  \widehat{X}^1_{\theta=0}  \\
                  \widehat{X}^2_{\theta=0} 
                      \end{array}
               \right), \quad
     R(\theta_{\psub,\qsub}) 
             \equiv
                 \left( \begin{array}{rr}
                     \mbox{cos}\,\theta_{\psub,\qsub} & 
                     \mbox{sin}\,\theta_{\psub,\qsub} \\ 
                    -\mbox{sin}\,\theta_{\psub,\qsub} & 
                     \mbox{cos}\,\theta_{\psub,\qsub} 
                        \end{array}
                 \right),
\label{eq:oscrotation}
\eeq
\beq
\mbox{sin}\, \theta_{\psub,\qsub}
            =\frac{\q \sqrt{G}/\sqrt{G_{11}}}
                  {\sqrt{G_{11}\p^2+ 2G_{12}\p\q
                  +G_{22}\q^2}},\quad
\mbox{cos}\, \theta_{\psub,\qsub}
            =\frac{\p \sqrt{G_{11}}
                  +\q G_{12}/\sqrt{G_{11}}}
                  {\sqrt{G_{11}\p^2+ 2G_{12}\p\q
                  +G_{22}\q^2}}.
\label{eq:sincos}
\eeq 
Then, the momentum is quantized in units of the 
inverse of the $(\p,\q)$ cycle's length for the Neumann  
direction,
while the length of the open string is quantized 
in units of the distance between the nearest 
trajectories for the Dirichlet 
direction:
\beqa
& &\widehat{p}_{\theta=\theta_{\ptiny,\qtiny}}^1 
        = \frac{m}{\sqrt{(G_{11}\p^2
          + 2G_{12}\p\q+G_{22}\q^2)\alpha'}}, \non \\
& &\widehat{X}_{\theta=\theta_{\ptiny,\qtiny}}^2(\tau, \pi)
        = \widehat{X}^2_{\theta=\theta_{\ptiny,\qtiny}}(\tau,0)
          + \frac{n \,\, 2 \pi \sqrt{\alpha'}\sqrt{G}}
              {\sqrt{G_{11}\p^2
               + 2G_{12}\p\q+G_{22}\q^2}},
\eeqa
$(m,n \in \ZZ)$ where 
\beq
\widehat{p}_{\theta=\theta_{\ptiny,\qtiny}}^a
= \frac{1}{2\pi \alpha'}\int_0^{\pi} d\sigma
[ \partial_{\tau}\widehat{X}_{\theta=\theta_{\ptiny,\qtiny}}^a
+(R(\theta_{\psub,\qsub})\widehat{B}
 R(\theta_{\psub,\qsub})^{-1})_{ab} \,
 \partial_{\sigma}\widehat{X}_
                           {\theta=\theta_{\ptiny,\qtiny}}^b ],
\eeq 
with
$\widehat{B}_{ab}=e^{\mu}_{\, \, a}
e^{\nu}_{\, \, b}B_{\mu \nu}$.
Solving these quantization conditions under
the boundary conditions (\ref{eq:openbc}), 
we get the following expansions 
\beqa
 \widehat{X}_{\theta=\theta_{\ptiny,\qtiny}}^1
 &=&  \frac{(m-B_{12}n) \,\, 2 \sqrt{\alpha'}}
              {\sqrt{G_{11}\p^2
               + 2G_{12}\p\q+G_{22}\q^2}}\,\, \tau
     +(\mbox{the oscillator part}),          \non \\
 \widehat{X}_{\theta=\theta_{\ptiny,\qtiny}}^2
 &=&  \frac{n \,\, 2 \sqrt{\alpha'}\sqrt{G}}
              {\sqrt{G_{11}\p^2
               + 2G_{12}\p\q+G_{22}\q^2}}\,\, \sigma
     +(\mbox{the oscillator part}).
\eeqa
Substituting them into 
the momentum-winding part of the open-string hamiltonian
and taking the summation, we have (\ref{eq:momwindo}). 
\vspace{0.3cm} \\
{\it The closed-string channel} \vspace{0.2cm}

In the closed-string channel, the partition function
is defined as the tree amplitude by using the boundary
state: 
\beq
 \hspace{0.5cm}<_{\,}^{\hspace{-0.7cm}\mbox{\scriptsize{tot}}}
  B|\Delta |B>^{\mbox{\scriptsize{tot}}}
= \frac{\alpha' \pi}{2}\int_0^{\infty}\frac{dt}{t^2}
  \hspace{0.5cm}<_{\,}^{\hspace{-0.7cm}\mbox{\scriptsize{tot}}}
  B|\tilde{q}^{\frac{H^{(\mbox{\tiny{c}})}}{2}}
  |B>^{\mbox{\scriptsize{tot}}},
\label{eq:closedpf}
\eeq
where $\Delta$ and  
$H^{(\mbox{\scriptsize{c}})}$ are 
the propagator and the hamiltonian of the closed string.
The boundary state $|B>^{\mbox{\scriptsize{tot}}}$
is a tensor product of the oscillator part 
and the zero-mode part;
the contributions from the compact directions
as well as the overall normalization
depend on $(\p,\q)$.
They are determined 
by solving the geometric boundary conditions and 
demanding that the above amplitude is equal to
$Z^{\mbox{\scriptsize{tot}}}_{\theta=\theta_{\ptiny,\qtiny}}$
(open-closed correspondence).
This is a generalization of, {\it e.g.}, 
\cite{GS,AK} to arbitrary $(\p,\q)$
branes on general tori.

The oscillator part of the boundary state 
is defined (\cite{DiVecchiarev}, for example) 
so that it satisfies the corresponding geometric
boundary conditions and provides the supersymmetric amplitude.
We refer to this state for 
the $(\p,\q)$ brane as
$|\theta_{\psub,\qsub}>_{\mbox{\scriptsize{osc}}}$
and normalize such that
\beq
 <^{\,}_{\hspace{-0.7cm}\mbox{\scriptsize osc}}
 \theta_{\psub,\qsub}|
           \tilde{q}^{\frac{H^{(\mbox{\tiny{c}})}}{2}}
                         |\theta_{\psub,\qsub}>
                         _{\mbox{\scriptsize osc}}
 = \frac{1}{\eta^6(\tilde{\tau})}\cdot
   \frac{1}{\eta^2(\tilde{\tau})}\cdot
   \frac{1}{2}\frac{\vartheta_3^4(\tilde{\tau},0)
                   -\vartheta_4^4(\tilde{\tau},0)
                   -\vartheta_2^4(\tilde{\tau},0)}
                   {\eta^4(\tilde{\tau})},
\label{eq:inner}
\eeq
where the each factor comes from the noncompact bosons,
the compact bosons and the fermions, respectively.    
The $\theta_{\psub,\qsub}$ dependences (\ref{eq:oscrotation})
of the compact-boson oscillators 
in the bra and the ket states cancel out.

The zero-mode parts of the boundary state are as follows.
For the noncompact directions, the boundary state
$|\vec{k},\vec{0}>$ has nonzero momenta $\vec{k}$ 
(zero momenta $\vec{0}$) for the Dirichlet (Neumann) directions,
and is normalized as \cite{DiVecchia}
\beq
 <\vec{k},\vec{0}|\vec{k'},\vec{0}>=V_p (2 \pi)^{8-p} 
                                    \delta^{(8-p)}
                                    (\vec{k}-\vec{k'}).
\label{eq:momnorm}
\eeq
For the torus directions,
in terms of the reference local Lorentz coordinates  
\beqa
& &\widehat{X}_{\theta=0}^a = \widehat{x}^a +\frac{\alpha'}{2}
           \widehat{p}^a_{\mbox{\scriptsize{R}}}(\tau -\sigma)
         + \frac{\alpha'}{2}
           \widehat{p}^a_{\mbox{\scriptsize{L}}}(\tau +\sigma)
                + (\mbox{the oscillator part})\, ,   \non   \\
& &\widehat{p}^a_{\mbox{\scriptsize{L}}}=
          (n_{\mu}+m^{\nu}(B_{\nu \mu}
                               +G_{\nu \mu}))
          \frac{e^{\mu}_{\, a}}{\sqrt{\alpha'}}\, , \quad
\widehat{p}^a_{\mbox{\scriptsize{R}}}=
          (n_{\mu}+m^{\nu}(B_{\nu \mu}
                               -G_{\nu \mu}))
          \frac{e^{\mu}_{\, a}}{\sqrt{\alpha'}},
\label{eq:momwind}       
\eeqa
the boundary conditions are written as 
\beqa
& &(\mbox{cos}\, \theta_{\psub,\qsub} \,
                 (\widehat{p}_{\mbox{\scriptsize{L}}}^1
                 +\widehat{p}_{\mbox{\scriptsize{R}}}^1) 
    +\mbox{sin}\, \theta_{\psub,\qsub} \,
                 (\widehat{p}_{\mbox{\scriptsize{L}}}^2
                 +\widehat{p}_{\mbox{\scriptsize{R}}}^2))|B>=0,
                                                    \non  \\
& & (-\mbox{sin}\, \theta_{\psub,\qsub} \,
                 (\widehat{p}_{\mbox{\scriptsize{L}}}^1
                 -\widehat{p}_{\mbox{\scriptsize{R}}}^1) 
    +\mbox{cos}\, \theta_{\psub,\qsub} \,
                 (\widehat{p}_{\mbox{\scriptsize{L}}}^2
                 -\widehat{p}_{\mbox{\scriptsize{R}}}^2))|B>=0,
\eeqa
which are equivalent to
\beq
(\q m^1-\p m^2)|B>=0, \quad 
(\p n_1+\q n_2)|B>=0.
\eeq
Plugging the solution
$(m^1,m^2)=(m \p,m \q)$, $(n_1,n_2)=(-n\q, n\p)$
($m,n \in \ZZ$) back into (\ref{eq:momwind}), we obtain
the two-dimensional lattice $\Lambda_{\psub,\qsub}$: 
\beqa
 \widehat{p}^1_{\mbox{\scriptsize{L}}}
    &=&  \frac{1}{\sqrt{\alpha' G_{11}}} 
         [-n\q+m\{ \p G_{11}+\q(G_{12}-B_{12})\}],
                                                     \non \\
 \widehat{p}^2_{\mbox{\scriptsize{L}}}
     &=& \frac{1}{\sqrt{\alpha' G_{11}G}}
         [n\{ \q G_{12}+\p G_{11}\}
                +m\{ \p G_{11}B_{12}
                     +\q(G+G_{12}B_{12})\}],       \non \\
 \widehat{p}^1_{\mbox{\scriptsize{R}}}
    &=&  \frac{1}{\sqrt{\alpha' G_{11}}} 
         [-n\q-m\{ \p G_{11}+\q(G_{12}+B_{12})\}],     
 \label{eq:momwindlat}                                  \\
 \widehat{p}^2_{\mbox{\scriptsize{R}}}
    &=&  \frac{1}{\sqrt{\alpha' G_{11}G}}
         [n\{ \q G_{12}+\p G_{11}\}
                +m\{ \p G_{11}B_{12}
                    -\q(G-G_{12}B_{12})\}].            \non
\eeqa
Thus, the momentum-winding part of
the boundary state can be written as
\beq
|\theta_{\psub,\qsub}>_0= 
     \sum_{\Lambda_{\ptiny, \qtiny}}
     |\widehat{p}^1_{\mbox{\scriptsize{L}}},
      \widehat{p}^2_{\mbox{\scriptsize{L}}};
      \widehat{p}^1_{\mbox{\scriptsize{R}}},
      \widehat{p}^2_{\mbox{\scriptsize{R}}}>.
\vspace{-0.2cm}
\eeq 
By evaluating the corresponding part of the
hamiltonian $H_0^{(\mbox{\scriptsize{c}})}$
in the amplitude, we obtain 
\beqa
  Z^{\mbox{\scriptsize{closed}}}
    _{\theta=\theta_{\ptiny,\qtiny},0}
    (\tilde{\tau})
    &\equiv& \,_0<\theta_{\psub,\qsub}           
          |\tilde{q}^{\frac{ H^{(\mbox{\tiny{c}})}_{0}}{2}}
          |\theta_{\psub,\qsub}>_0             \non \\   
    &=& \sum_{n,m \in \ZZsub}
      \tilde{q}^{\frac{G_{11}\psub^2
                      +2G_{12}\psub\qsub+G_{22}\qsub^2}
                      {4 G}
                (n^2+2 B_{12} nm +(G+B_{12}^2)m^2)},
\label{eq:closedpfpq}
\eeqa
which, in fact, is related to the open-string one
(\ref{eq:momwindo}) via the Poisson resummation:
\beq
   \frac{1}{\eta^2(\tau)}\,Z^{\mbox{\scriptsize{open}}}
      _{\theta=\theta_{\ptiny,\qtiny},0}(\tau)
=  C_{\psub,\qsub}\,
   \frac{1}{\eta^2(\tilde{\tau})}\,
   Z^{\mbox{\scriptsize{closed}}}
    _{\theta=\theta_{\ptiny,\qtiny},0}(\tilde{\tau})\, ,
\quad
   C_{\psub,\qsub}
     = \frac{G_{11}\p^2+2G_{12}\p\q+G_{22}\q^2}
            {2\sqrt{G}}.             \vspace{0.2cm}
\label{eq:Poisson}
\eeq     
{\it The open-closed correspondence} \vspace{0.2cm}

The total boundary state
which satisfies the open-closed correspondence
is then obtained by
\beq
 |\theta=\theta_{\psub,\qsub}>^{\mbox{\scriptsize{tot}}}
 = \frac{N_p}{2} \sqrt{C_{\psub,\qsub}}
   \int(\frac{dk}{2\pi})^{8-p}
   |\theta_{\psub,\qsub}>_{\mbox{\scriptsize{osc}}}\times
   |\theta_{\psub,\qsub}>_{0}\times
   |\vec{k},\vec{0}>,
\label{eq:totbs}
\eeq
where
\beq
N_p=\sqrt{2}\, T_p , \quad
T_p\equiv 2^{3-p}\pi^{7/2-p}\alpha'^{(3-p)/2}.
\label{eq:nptp}
\eeq
In fact, by using (\ref{eq:inner}), 
(\ref{eq:momnorm}), (\ref{eq:closedpfpq}) and 
(\ref{eq:Poisson}),
it is easy to show that
\beq
  \hspace{0.5cm}<_{\,}^{\hspace{-0.7cm}\mbox{\scriptsize{tot}}}
  \theta=\theta_{\psub,\qsub}|
  \Delta
  |\theta=\theta_{\psub,\qsub}>^{\mbox{\scriptsize{tot}}}
 \,\, \, = \,\,\,
  Z^{\mbox{\scriptsize{tot}}}_{\theta=\theta_{\ptiny,\qtiny}}.
\eeq

Let us discuss the meaning of the prefactor. 
It is known 
that $T_p$ in (\ref{eq:nptp}) 
is the normalization factor of the D$p$-brane boundary state 
in ten-dimensional Minkowski space,
and is related to its tension as \cite{DiVecchia:1997pr}
\beq
\tau_p= \frac{1}{\kappa_{10}}T_p
\eeq
($\kappa_{10}$ is 
the ten-dimensional gravitational constant).
After compactification on a torus, the gravitational
constant reads
\beq 
\frac{1}{\kappa_8}
  =\frac{1}{\kappa_{10}}\sqrt{V_{\mbox{\scriptsize{torus}}}}\, ,
\quad
V_{\mbox{\scriptsize{torus}}}=\sqrt{G}(2\pi \sqrt{\alpha'})^2.
\eeq
Thus, the tension $\tau_{\psub,\qsub}$ of the $(\p,\q)$ brane 
in eight dimensions becomes 
\beq
 \tau_{\psub,\qsub}
   =\frac{1}{\kappa_{8}}N_p \sqrt{C_{\psub,\qsub}}
   =\tau_{p}\,\, 2\pi\sqrt{\alpha'
           (G_{11}\p^2+2G_{12}\p\q+G_{22}\q^2)},
\eeq   
which is consistent with the geometric picture.
\subsection{Branes at an angle}
~~~
In this subsection, we consider the
intersection of the $(\tilde{\p},\tilde{\q})$ brane
with the $(\p,\q)$ brane at an angle.
The one-loop partition function has been obtained
\cite{Bachas,Arfaei} for the
open strings ending on such  
branes in a noncompact target space.
In this case, the open-string modes
receive angle-dependent twists, 
and hence no zeromode exists, in particular.
Therefore, even when we consider branes 
intersecting in a compact space,
there is no modification of the partition function by
either the momentum quantization or the appearance of the
winding modes.
However, the volume integral in the trace in (\ref{eq:openpf})
receives the change because
the intersection number of the two branes is 1
for the noncompact case, whereas
$(\tilde{\p},\tilde{\q})\cdot (\p,\q)$
$\equiv \tilde{\q}\p-\tilde{\p}\q$ for the compact case. 
We thus multiply (the absolute value of) this extra factor
to the total partition function
and obtain
\beqa
  Z^{\mbox{\scriptsize{tot}}}
    _{\theta_{\tilde{\ptiny},\tilde{\qtiny}},\,
      \theta_{\ptiny,\qtiny}}
&=& 2(\tilde{\p},\tilde{\q})\cdot (\p,\q)V_p 
    \int_0^{\infty}\frac{dt}{2t}
       (8\pi^2t \alpha')^{-\frac{p}{2}}
       \frac{1}{\eta^6(\tau)}                \non  \\
& & \hspace{-0.6cm}\cdot \frac{1}{2}
         \frac{\vartheta_3^3(\tau,0)
               \vartheta_3(\tau,
                   \frac{\Delta \theta}{\pi}\tau)
              -\vartheta_2^3(\tau,0)
               \vartheta_2(\tau,
                   \frac{\Delta \theta}{\pi}\tau)
              -\vartheta_4^3(\tau,0)
               \vartheta_4(\tau,
                   \frac{\Delta \theta}{\pi}\tau)}
              {\eta^{3}(\tau)
               (-i\vartheta_1(\tau,
                   \frac{\Delta \theta}{\pi}\tau))},
\label{eq:openpfangle}
\eeqa 
where $\Delta \theta
=\theta_{\tilde{\psub},\tilde{\qsub}}-\theta_{\psub,\qsub}$. 
The sign of $\mbox{sin}(\Delta \theta \tau)$
in the denominator and that of the intersection number 
in the numerator cancel out.

We can alternatively {\it derive} this partition function
as the amplitude between the boundary states (\ref{eq:totbs})
with different angles.
The amplitude for the oscillator part is 
\beqa
 <^{\,}_{\hspace{-0.7cm}\mbox{\scriptsize osc}}
 \theta_{\tilde{\psub},\tilde{\qsub}}|
           \tilde{q}^{\frac{H^{(\mbox{\tiny{c}})}}{2}}
                         |\theta_{\psub,\qsub}>
                         _{\mbox{\scriptsize osc}}
 &=& \frac{1}{\eta^6(\tilde{\tau})}\cdot
  \frac{2\mbox{sin}\,\Delta \theta\,\eta(\tilde{\tau})}
       {\vartheta_1(\tilde{\tau},\frac{\Delta \theta}{\pi})}\non \\
 & &\hspace{-2cm} 
  \cdot \frac{1}{2}\frac{\vartheta_3^3(\tilde{\tau},0)
                    \vartheta_3(\tilde{\tau},
                                \frac{\Delta \theta}{\pi})
                   -\vartheta_4^3(\tilde{\tau},0) 
                    \vartheta_4(\tilde{\tau},
                                \frac{\Delta \theta}{\pi})
                   -\vartheta_2^3(\tilde{\tau},0)
                    \vartheta_2(\tilde{\tau},
                                \frac{\Delta \theta}{\pi})}
                   {\eta^4(\tilde{\tau})},
\label{eq:innerpq}
\eeqa
where the states are normalized as in (\ref{eq:inner}).
The $\Delta \theta$ dependence arises 
because the oscillators of the compact bosons
(\ref{eq:oscrotation}) and their superpartners 
(including the RR zero modes
\footnote{$\mbox{cos}\,\Delta \theta$ 
in $\vartheta_2(\tilde{\tau},\frac{\Delta \theta}{\pi})$
comes from the RR vacuum amplitude.})
in the bra and the ket states
are relatively rotated.
For the momentum-winding part, (\ref{eq:closedpfpq}) 
is replaced by 
\beq
<^{\,}_{\hspace{-0.5cm}0}\theta_{\tilde{\psub},\tilde{\qsub}}|
      \tilde{q}^{\frac{H_0^{(\mbox{\tiny{c}})}}{2}}
      |\theta_{\psub,\qsub}>_0 \,\,\, = \,\,1.
\label{eq:momwindangle}
\eeq
This holds because the momenta (\ref{eq:momwindlat})
between the two boundary states coincide
only at one lattice point,
$\widehat{p}^a_L=\widehat{p}^a_R=0$.
Another identity we use is
\beq
\sqrt{C_{\tilde{\psub},\tilde{\qsub}}}
\sqrt{C_{\psub,\qsub}}\, 2\,\mbox{sin}\,\Delta \theta
=(\tilde{\p},\tilde{\q})\cdot (\p,\q).
\eeq
Then, we find 
\beq
  \hspace{0.5cm}<_{\,}^{\hspace{-0.7cm}\mbox{\scriptsize{tot}}}
  \theta=\theta_{\tilde{\psub},\tilde{\qsub}}|
  \Delta
  |\theta=\theta_{\psub,\qsub}>^{\mbox{\scriptsize{tot}}}
 \,\,\,\,=\,\,\,
   Z^{\mbox{\scriptsize{tot}}}
    _{\theta_{\tilde{\ptiny},\tilde{\qtiny}},\,
      \theta_{\ptiny,\qtiny}}
\eeq
as expected.
This also implies that the boundary
states (\ref{eq:totbs}) are mutually consistent.
\subsection{Partition functions normalized in terms of CFT}
~~~
In later sections, we will construct boundary states 
in the $(k=1)^3$ and $(k=2)^2$ Gepner models so that they
reproduce the geometric partition functions 
for the $SU(3)$ and $SU(2)^2$ torus, respectively.
For this purpose, it is convenient to extract the
`CFT piece' from (\ref{eq:openpfpq}) and (\ref{eq:openpfangle})
by discarding the {\it common} kinematical factors 
(and $1/\eta^6(\tau)$) as
\beqa
& & \hspace{-1cm}
    Z^{\mbox{\scriptsize{open}}}
    _{\theta=\theta_{\ptiny,\qtiny}}(\tau)
    \equiv \frac{1}{\eta^2(\tau)}\,
              Z^{\mbox{\scriptsize{open}}}
               _{\theta=\theta_{\ptiny,\qtiny},0}(\tau)\,
              \frac{1}{2}\frac{\vartheta_3^4(\tau,0)
                              -\vartheta_2^4(\tau,0)
                              -\vartheta_4^4(\tau,0)}
                              {\eta^{4}(\tau)},
\label{eq:openpfpqcft}                       \\
& & \hspace{-1cm} 
    Z^{\mbox{\scriptsize{open}}}
    _{\theta_{\tilde{\ptiny},\tilde{\qtiny}},\,
      \theta_{\ptiny,\qtiny}}(\tau)
    \equiv (\tilde{\p},\tilde{\q})\cdot (\p,\q)         \non \\
&  & \hspace{1cm}  \cdot  \frac{1}{2}
         \frac{\vartheta_3^3(\tau,0)
               \vartheta_3(\tau,
                   \frac{\Delta \theta}{\pi}\tau)
              -\vartheta_2^3(\tau,0)
               \vartheta_2(\tau,
                   \frac{\Delta \theta}{\pi}\tau)
              -\vartheta_4^3(\tau,0)
               \vartheta_4(\tau,
                   \frac{\Delta \theta}{\pi}\tau)}
              {\eta^{3}(\tau)\,
               (-i \vartheta_1(\tau,
                   \frac{\Delta \theta}{\pi}\tau))}.
\label{eq:openpfanglecft}
\eeqa

\setcounter{equation}{0}
\section{D1-branes in the $(k=1)^3$ model}
\subsection{Partition functions}
~~~
Let us now concentrate on the D1-branes
\footnote{Again, no spacelike noncompact Neumann directions are assumed 
below.}
in the $(k=1)^3$ model. 
As mentioned before, it describes the compactification on the special
torus, with backgrounds
\begin{eqnarray}
G_{\mu \nu}=\left(
\begin{array}{cc}1&\frac12\\ \frac12&1
\end{array}\right),~~~
B_{\mu \nu}=\left(
\begin{array}{cr}0&\frac12\\ -\frac12&0
\end{array}\right)
\end{eqnarray}
in our convention.
The Recknagel-Schomerus states are 
\footnote{Since $\chi^{k-l,s+2}_{m+k+2}=\chi^{l,s}_m$, the same 
state appears $2^3=8$ times in the summation; they all have the
same phase factor if the integer labels $(L_j,M_j,S_j)$ satisfy 
\mbox{$L_j+M_j+S_j=$even}. $|\alpha>_{\A}$ itself vanishes otherwise. }
\footnote{To be precise, this $B^{\lambda,\mu}_\alpha$ (\ref{B}) is 
the phase-factor piece of (\ref{B^lambdamu_alpha}) and the remaining real 
constant is absorbed into the prefactor of (\ref{|alpha>}). But still we
use the same symbol 
for notational simplicity.}
\begin{eqnarray}
|\alpha>_{\A}&=&\frac1{2^{\frac52} 3^{\frac14}}
\sum_{s_0,l_j,m_j,s_j}^\beta
B^{\lambda,\mu}_\alpha
|s_0\!>\!\!>
\prod_{j=1}^3|l_j,m_j,s_j\!>\!\!>_{\A},
\label{|alpha>}
\\
\nonumber\\
B^{\lambda,\mu}_\alpha&=&
e^{\pi i \left[\frac{s_0^2}2+\sum_{j=1}^3l_jL_j
+\frac{s_0S_0}2+\sum_{j=1}^3 (
\frac{m_jM_j}3-\frac{s_jS_j}2
)
\right]}. \label{B}
\end{eqnarray}
As we have seen in section 2, the partition function for a 
boundary state $|\alpha>_{\A}$ with itself can be written as 
a sum of $\beta$-orbits. This motivates us to introduce 
the notation 
\begin{eqnarray}
\mbox{Orbit}(s_0;(l_j,m_j,s_j);z_j)\hspace{80mm}{}
\nonumber\\
\equiv\sum_{\nu_0\in \ZZsub_{12}}
(-1)^{\nu_0}\chi^{SO(6)}_{s_0+\nu_0}(\tau)
\chi^{l_1,s_1+\nu_0}_{m_1+\nu_0}(\tau,z_1)
\chi^{l_2,s_2+\nu_0}_{m_2+\nu_0}(\tau,z_2)
\chi^{l_3,s_3+\nu_0}_{m_3+\nu_0}(\tau,z_3).
\end{eqnarray}
The minimal characters are given by 
\begin{eqnarray}
\chi^{l,s}_{m}(\tau,z)=\frac{\delta^{(2)}_{l+m+s,0}}{\eta(\tau)}
\Theta_{2m-3s,6}(\tau,\frac z3).
\end{eqnarray}
The $SO(6)$ characters are
\begin{eqnarray}
\chi^{SO(6)}_{s_0}(\tau)&=&\frac1{\eta^3(\tau)}
\left[
(\Theta_{-s_0,2}(\tau,0))^3
+3\Theta_{-s_0,2}(\tau,0)(\Theta_{-s_0+2,2}(\tau,0))^2
\right].
\end{eqnarray}
$\mbox{Orbit}(s_0;(l_j,m_j,s_j);z_j)$ identically vanishes if 
the  
$\beta$-condition is satisfied \cite{Gepner}:
\begin{eqnarray}
\frac{s_0}4+\frac12\sum_{j=1}^3
\left(\frac{m_j}3-\frac{s_j}2\right)
=\frac12~~~~~~(\mbox{mod \ZZ})
\end{eqnarray}
and if
\begin{eqnarray}
z_1+z_2+z_3=0.
\end{eqnarray}
Indeed, one can then write
\begin{eqnarray}
\mbox{Orbit}(s_0;(l_j,m_j,s_j);z_j)
&=&\sum_{j\in \ZZsub_3}\sum_{\nu\in \ZZsub_4}\frac{(-1)^\nu}{\eta^3(\tau)}
\chi^{SO(6)}_{s_0-\nu}(\tau)
\nonumber\\
&&\cdot
\Theta_{\widehat{m}_1+4j+\nu,6}(\tau, \frac{z_1}{3})
\Theta_{\widehat{m}_2+4j+\nu,6}(\tau, \frac{z_2}{3})
\Theta_{\widehat{m}_3+4j+\nu,6}(\tau, \frac{z_3}{3})\nonumber\\
&=&
(-1)^{s_0}(\mbox{Jacobi})
\cdot\frac1{\eta^2(\tau)}
\sum_{s\in \ZZsub_2}
\Theta_{12s+\widehat{m}_2-\widehat{m}_1,12}\left(
\tau,\frac{z_2-z_1}6\right)\nonumber\\
&&~~~~~~~~~~~~~~~~~~~~~\cdot\Theta_{
        4s-\frac13(2\widehat{m}_3-\widehat{m}_1-\widehat{m}_2),4}
\left(\tau,\frac{z_1+z_2}2\right),\label{Orbit}
\end{eqnarray}
where
\begin{eqnarray}
(\mbox{Jacobi})&\equiv&
\sum_{\nu\in \ZZsub_4}
\frac{(-1)^\nu}{\eta(\tau)}\chi^{SO(6)}_{-\nu}(\tau)
\Theta_{\nu+2,2}(\tau,0)\nonumber\\
&=&\frac1{2\eta^4(\tau)}\left(
\vartheta_3^4(\tau)-\vartheta_4^4(\tau)-\vartheta_2^4(\tau)\right)
=0,
\end{eqnarray}
\begin{eqnarray}
\widehat{m}_j\equiv2m_j-3s_j~~~~~~~~~~~~(j=1,2,3).
\end{eqnarray}
In the last line (\ref{Orbit}), we have used the 
multiplication formula for the theta functions.
%
%
The $l_j$ dependences are only through 
$\delta^{(2)}_{l_j+m_j+s_j,0}$, and they are omitted 
in (\ref{Orbit}).

With this notation, the partition function for the boundary state
$|\alpha>_{\A}$ (\ref{Z_alphaalpha}) with itself can be written as
\begin{eqnarray}
Z^A_{\alpha\alpha}(\tau)&=&
\sum_{\nu_1,\nu_2,\nu_3\in \ZZsub_2}
\mbox{Orbit}\left(2-2\sum_{j=1}^3 \nu_j
;(0,0,2\nu_j);0\right)
\nonumber\\
&=&2(\mbox{Jacobi})
\cdot\frac1{\eta^2(\tau)}\sum_{m,n\in \ZZsub}
q^{m^2+mn+n^2},
\end{eqnarray}
which
\footnote{A factor of 2 comes from the fact that only a half of 
the choices of $\nu_j\in \ZZsmall_2$ represent independent 
$\beta$-orbits.}
precisely coincides with 
$2Z^{\mbox{\scriptsize open}}_{\theta=0}(\tau)$,
the open-string partition function 
(\ref{eq:momwindo})\,(\ref{eq:openpfpqcft}) for $\theta=0$. 
Thus we have 
\begin{eqnarray}
|\theta=0>&=&\frac1{\sqrt{2}}|\alpha>_{\A},
\end{eqnarray}
where $\alpha=(S_0,(L_j,M_j,S_j))$
are a set of arbitrary integer labels satisfying 
$L_j+M_j+S_j=\mbox{even}$.

Let us next express 
\begin{eqnarray}
Z_{\theta=\theta_{\ptiny,\qtiny}}^{\rm open}
(\tau)&=&
(\mbox{Jacobi})
\cdot\frac1{\eta^2(\tau)}
\sum_{m,n\in \ZZsub}
q^{\frac{m^2+mn+n^2}{\mbox{\scriptsize\boldmath $p$}^2
+\mbox{\scriptsize\boldmath $pq$}+\mbox{\scriptsize\boldmath $q$}^2}}
\label{Z_thetapq}
\end{eqnarray}
for a general pair of relatively prime integers 
$\p,\q$.  
The angle $\theta_{\psub,\qsub}$ (\ref{eq:sincos})
of the brane is given by
\begin{equation}
\sin\theta_{\psub,\qsub}
=\frac{\frac{\sqrt{3}}2 \q}{\sqrt{\p^2+\p\q+\q^2}},
~~~\cos\theta_{\psub,\qsub}=
\frac{\p+\frac\q2 }{\sqrt{\p^2+\p\q+\q^2}}
\end{equation}
in the present case.
We make use of the equation
\begin{eqnarray}
&&\sum_{m,n\in \ZZsub}
q^{\frac{m^2+mn+n^2}{
\mbox{\scriptsize\boldmath $p$}^2
+\mbox{\scriptsize\boldmath $pq$}
+\mbox{\scriptsize\boldmath $q$}^2}}
=\sum_{\Delta m\in \ZZsub_{\ptiny^2+\ptiny \qtiny +\qtiny^2}} 
q^{\frac{(\Delta m)^2}{\common}}
\!\!\!\sum_{s,\nu_1,\nu_2\in \ZZsub_2}
\Theta_{12s+6\nu_1,12}
\left(\tau,
\frac{(\p+\q)
      \tau\Delta m}
{2(\p^2+\p\q+\q^2)}
\right)~~~~\nonumber\\
&&~~~~~~~~~~~~~~~~~~~~~~~~~~~~~~~~~~~~~~~~~~~~~~~~~~~~~\cdot
\Theta_{4s+2\nu_1+4\nu_2,4}
\left(\tau,
\frac{(\p-\q)
      \tau\Delta m}
{2(\p^2+\p\q+\q^2)}
\right).
\end{eqnarray}
Plugging this equation into (\ref{Z_thetapq}) and comparing 
with (\ref{Orbit}), we find 
\begin{eqnarray}
Z_{\theta=\theta_{\ptiny,\qtiny}}^{\rm open}
(\tau)
&=&\sum_{\Delta m\in \ZZsub_{\ptiny^2+\ptiny \qtiny+\qtiny^2}} 
q^{\frac{(\Delta m)^2}{\common}}\nonumber\\
&&\cdot\frac12\sum_{\nu_1,\nu_2,\nu_3\in \ZZsub_2}
\mbox{Orbit}\left(2-2\sum_{j=1}^3\nu_j;(0,0,2\nu_j);\tau z_j(\Delta m)
\right)
\label{Z_pq}
\end{eqnarray}
with
\begin{eqnarray}
z_1(\Delta m)&=&\frac{-\p-2\q}{\p^2+\p\q+\q^2}\Delta m,
\nonumber\\
z_2(\Delta m)&=&\frac{2\p+\q}{\p^2+\p\q+\q^2}\Delta m,
\label{zjDeltam}\\
z_3(\Delta m)&=&\frac{-\p+\q}{\p^2+\p\q+\q^2}\Delta m.
\nonumber
\end{eqnarray}
$Z_{\theta=\theta_{\ptiny,\qtiny}}^{\rm open}
(\tau)$ can thus be obtained by first shifting $z_j$ of each minimal 
character in $Z_{\theta=0}^{\rm open}(\tau)$
by a fractional unit times $\tau\Delta m$,  
and then summing over $\Delta m\in \ZZ_{\common}$. 
Such a shift of $z$ linear in $\tau$ is known as a `spectral flow'.
Note that the $\beta$-orbit itself is also 
a spectral-flow orbit of the total $U(1)$ current. These two kinds of 
spectral flows are orthogonal to each other since 
\begin{eqnarray}
z_1(\Delta m)+z_2(\Delta m)+z_3(\Delta m)=0.
\end{eqnarray}

\subsection{Ishibashi states in different realizations}
\label{Istates}
~~~
To construct boundary states reproducing the partition functions
(\ref{Z_pq}), we will first generalize the ordinary $N=2$ Ishibashi
states to a certain one-parameter family of them. We restrict ourselves
to the $k=1$ case. 
The A-type boundary states $|B>\!\!>_{\A}$
are defined to be the states such that 
\begin{eqnarray}
(L_n-\overline{L}_{-n})|B>\!\!>_{\A}=
(J_n-\overline{J}_{-n})|B>\!\!>_{\A}=
(G_r^\pm+i\eta\overline{G}_{-r}^\mp)|B>\!\!>_{\A}=0
\label{A-conditions}
\end{eqnarray}
$(\eta=\pm1)$. If $k=1$, the $N=2$ superconformal currents
are realized by a single free boson $\phi(z)$ as
\begin{eqnarray}
T(z)=-\frac12(\partial\phi)^2,
~~~J(z)=\frac i{\sqrt{3}}\partial\phi,
~~~G^\pm(z)=\sqrt\frac23 e^{\pm i\sqrt{3}\phi(z)}
\label{realization}
\end{eqnarray}
(and similar expressions obtained by the replacement 
$\phi(z)\rightarrow\overline{\phi}(\overline{z})$ for 
$\overline{T}(\overline{z})$, $\overline{J}(\overline{z})$
and $\overline{G}^\pm(\overline{z})$). Their mode expansions are 
\begin{eqnarray}
T(z)=\sum_{n\in \ZZsub} \frac{L_n}{z^{n+2}},
~~~J(z)=\sum_{n\in \ZZsub} \frac{J_n}{z^{n+1}},
~~~G^\pm(z)=\sum_{r\in \ZZsub+\nu} \frac{G^\pm_r}{z^{r+3/2}}
\end{eqnarray}
($\nu=\frac12$ or $0$ depending on NS or R) and
\begin{eqnarray}
\partial\phi(z)=\sum_{n\in \ZZsub} \frac{a_n}{z^{n+1}}.
\end{eqnarray}
The anti-holomorphic fields (with $\overline{\phantom{a}}$'s) 
are also expanded in a similar fashion.

The first two 
A-boundary conditions (\ref{A-conditions}) are equivalent to 
the Dirichlet boundary condition for the free boson 
$\phi+\overline{\phi}$
\begin{eqnarray}
(a_n-\overline{a}_{-n})|B>\!\!>_{\A}=0. \label{Dirichlet}
\end{eqnarray}
In general, an Ishibashi state is obtained by first gluing every 
state of an irreducible representation with its anti-linear 
transform, and then summing over all states in the whole 
representation space. Thus, an $N=2$ A-type Ishibashi state is 
not a single  Dirichlet boundary state of the free boson, 
but a sum of such states  
whose  momenta lie on the momentum lattice of the realization. 
The relative phase factor of each state with different momentum 
in the summation is determined by the last equation 
of (\ref{A-conditions}).

Let the boundary be at $|z|=1$. Writing the boundary values of 
$\phi(z)$, $\overline{\phi}(\overline{z})$ as functions of 
$\zeta=-i\log z$, the Dirichlet condition (\ref{Dirichlet}) 
translates into 
\begin{eqnarray}
\partial_\zeta(\phi+\overline{\phi})=0. 
\end{eqnarray}
The constant value of $\phi+\overline{\phi}\equiv x$ at the 
boundary is an undetermined integration constant, being a modulus 
of the free-boson boundary state. The $G^\pm$ boundary conditions 
in (\ref{A-conditions}) fix this modulus to be 
\begin{eqnarray}
x=0
\end{eqnarray}
since
\begin{eqnarray}
e^{\pm\sqrt{3}i\phi(\zeta)}
+\eta e^{\mp\sqrt{3}i\overline{\phi}(\zeta)}=0.
\end{eqnarray}
(Using the label $s$, $\eta=\pm 1$ is absorbed into the overall 
constant of the $N=2$ boundary states with definite `G-parity'.)
Therefore, defining the free-boson vacua as 
\begin{eqnarray}
&&|\gamma> ~\equiv~ e^{i\gamma(\phi(z)+\overline{\phi}(\overline{z}))|0>},
\nonumber\\
&&a_0|\gamma>~=~\overline{a}_0|\gamma>~=~\gamma|\gamma>,
\end{eqnarray}
an $N=2$ Ishibashi state can be written in a sum of free-boson 
Dirichlet states
\begin{eqnarray}
|~l,m,s>\!\!>_{\A}
&=&\sum_{n\in \ZZsub}\exp\left(
\sum_{k=1}^\infty\frac{a_{-k}\overline{a}_{-k}}k
\right)
|\gamma_{m,s}+2\sqrt{3} n>,
\end{eqnarray}
where
\begin{eqnarray}
\gamma_{m,s}&=&\sqrt{3}\left(\frac m3-\frac s2
\right)
\end{eqnarray}
is the momentum of the $N=2$ vacuum state.

We next turn on the modulus
$(\phi+\overline{\phi})|
_{\mbox{\scriptsize boundary}}=x\neq 0$. The $G^{\pm}$ 
boundary conditions then change into 
\begin{eqnarray}
e^{\pm\sqrt{3}i\phi(\zeta)}
+\eta ~e^{\pm\sqrt{3}ix} e^{\mp\sqrt{3}i\overline{\phi}(\zeta)}=0
&\Longleftrightarrow&
(G_r^\pm+i\eta e^{\pm\sqrt{3}ix}\overline{G}_{-r}^\mp)
|B>\!\!>_{\mbox{\scriptsize A},x}=0.
\label{A,x}
\end{eqnarray}
Thus, $|B>\!\!>_{\mbox{\scriptsize A},x}$ 
(whose defining relations are  
(\ref{A-conditions}) with the last conditions replaced by (\ref{A,x})) 
is {\it not} an A-boundary state in the realization 
(\ref{realization}). However, if we use, for instance 
\begin{eqnarray}
\phi'(z)=\phi(z),~~~~~~
\overline{\phi}'(\overline{z})=\overline{\phi}(\overline{z})-x 
\end{eqnarray}
for the new realization 
\begin{eqnarray}
G'^\pm(z)=\sqrt\frac23 e^{\pm i\sqrt{3}\phi'(z)},~~~~~~
\overline{G}'^\pm(\overline{z})=\sqrt\frac23 e^{\pm
i\sqrt{3}\overline{\phi}'(\overline{z})},
\end{eqnarray}
it is clear that the modes of $G'^\pm(z)$, 
$\overline{G}'^\pm(\overline{z})$ satisfy the usual A-boundary 
conditions. 

Let us now define a new one-parameter family of Ishibashi states 
\mbox{$|l,m,s;\nu >\!\! >_{\A}$} by the equation
\footnote{More precisely, it should read 
$|\cdots>\!\!>_{\mbox{\scriptsize A},x}$
since it satisfies the same boundary conditions as 
\mbox{$|B>\!\!>_{\mbox{\scriptsize A},x}$} does.}
\begin{eqnarray}
|~l,m,s~;\nu>\!\!>_{\A}&=&
\sum_{n\in \ZZsub}e^{
\sum_{k=1}^\infty\frac{a_{-k}\overline{a}_{-k}}k
}
e^{-ix(\gamma_{m,s}+2\sqrt{3} n)}
|\gamma_{m,s}+2\sqrt{3} n>,
\end{eqnarray}
where $\nu\in \RR$ is the modulus
\begin{eqnarray}
x&=&-\frac{2\pi \nu}{\sqrt{3}}.
\end{eqnarray}
One of the important effects of the modulus is that it causes a shift 
in the `$z$' argument of characters in the transition amplitude
\begin{eqnarray}
&&{}_{\A}\!<\!\!<\tilde{l},\tilde{m},\tilde{s}~;\tilde{\nu},
|\tilde{q}^{L^{N=2}_0-\frac{c^{N=2}}{24}}
|l,m,s~;\nu\!>\!\!>_{\A}\nonumber\\
&&~~~~~~~~~~~~~~~~~~~~~~~~~~~~=\left(
\delta_{\tilde{l},l}
\delta^{(2(k+2))}_{\tilde{m},m}
\delta^{(4)}_{\tilde{s},s}
+\delta_{\tilde{l},k-l}
\delta^{(2(k+2))}_{\tilde{m},m+k+2}
\delta^{(4)}_{\tilde{s},s+2}
\right)\chi^{l,s}_m(\tilde\tau,\nu-\tilde{\nu}),~~~~~~~~~
\label{N=2innerproduct}
\end{eqnarray}
which replaces the relation (\ref{product}). Clearly, the original
A-boundary  state \mbox{$|l,m,s>\!\! >_{\A}$} is a special case for which
$\nu$  is set equal to $0$. 

To summarize, $|l,m,s~;\nu\!>\!\!>_{\A}$ is obtained 
by giving a phase factor of $e^{2\pi i\nu Q}$ to every state in 
the Ishibashi state $|l,m,s\!>\!\!>_{\A}$, depending on the
$U(1)$ charge $Q$ of each state. 
$|l,m,s~;\nu\!>\!\!>_{\A}$ is an A-boundary state for the $N=2$ 
generators in a realization in which the free-boson modulus is
shifted by $\nu$. 

\subsection{Boundary states at general angles} 
~~~
We will now construct the boundary state for D1-branes 
tilted at an angle of 
$\theta_{\mbox{\psub,\qsub}}$ by using the one-parameter family 
of $N=2$ Ishibashi states we have constructed in the previous 
subsection. Suppose that we replace each  Ishibashi state 
\mbox{$|l_j,m_j,s_j\!>\!\!>_{\A}$} in
$|\alpha>_{\A}$  (\ref{RSboundarystates}) by one with a shifted 
modulus $|l_j,m_j,s_j~;z_j\!>\!\!>_{\A}$. Define
\begin{eqnarray}
|\alpha~;z\!>_{\A}&\equiv&\frac1{\kappa_\alpha^A}
\sum_{\lambda,\mu}^\beta
B^{\lambda,\mu}_\alpha|\lambda,\mu~;z>\!\!>_{\A},\nonumber\\
|\lambda,\mu~;z>\!\!>_{\A}&\equiv&
|s_0>\!\!>\prod_{j=1}^r|l_j,m_j,s_j~;z_j\!>\!\!>_{\A},
\end{eqnarray}
where the coefficients $B^{\lambda,\mu}_\alpha$ are the same 
as before (given by (\ref{B^lambdamu_alpha})).
Using the formula for the modular transformations (\ref{modular}),
a calculation similar to (\ref{Z_alphaalphatilde}) 
leads us to the expression
\begin{eqnarray}
Z^A_{(\tilde{\alpha};\tilde{z})(\alpha;z)}(\tau)
&\equiv&{}_{\A}\!<\Theta\tilde{\alpha};\tilde{z}
|~\tilde{q}^{L_0-\frac
c{24}}|~\alpha ;z>_{\A}
\nonumber\\
&=&\sum_{\lambda',\mu'}\sum_{\nu_0=0}^{K-1}\sum_{\nu_1\ldots,\nu_r=0,1}
(-1)^{\nu_0}\delta^{(4)}_{s_0',
								2+\tilde{S}_0-S_0-\nu_0-2\sum_{j=1}^r\nu_j}\nonumber\\
&&\cdot\prod_{j=1}^r\left(
N_{L_j\tilde{L}_j}^{l_j'}
\delta^{(2(k_j+2))}_{m_j',\tilde{M}_j-M_j-\nu_0}
\delta^{(4)}_{s_j',\tilde{S}_j-S_j-\nu_0+2\nu_j}
\right)\nonumber\\&&\cdot
q^{\sum_{j=1}^r\frac{k_j}{2(k_j+2)}(\tilde{z}_j-z_j)^2}
\chi^{\lambda',\mu'}(\tau,\tau(z-\tilde{z})),
\end{eqnarray}
where
\begin{eqnarray}
{}_{\A}\!\!<\!\!<\!\lambda,\mu~;\tilde{z}|
~\tilde{q}^{L_0-\frac c{24}}
|\lambda,\mu~;z>\!\!>_{\A}
&=&\chi^{\lambda,\mu}(\tilde\tau,z-\tilde{z})\nonumber\\
&\equiv&
\chi^{SO(d)}_{s_0}(\tilde\tau)
\prod_{j=1}^r\chi^{l_j,s_j}_{m_j}(\tilde\tau,z_j-\tilde{z}_j),
\label{<lambdamu;z>}
\end{eqnarray}
and
\begin{eqnarray}
\chi^{\lambda',\mu'}(\tau,\tau(z-\tilde{z}))
&=&
\chi^{SO(d)}_{s'_0}(\tau)
\prod_{j=1}^r\chi^{l'_j,s'_j}_{m'_j}(\tau,\tau(z_j-\tilde{z}_j)).
\label{chi^lambdamu}
\end{eqnarray}
Thus we see that each minimal character gets a 
spectral flow in the open channel because of the modulus shift.

We have already expressed the partition function 
$Z_{\theta=\theta_{\ptiny,\qtiny}}^{\rm open}
(\tau)$ (\ref{Z_thetapq}) in a sum over spectral-flowed 
$\beta$-orbits (\ref{Z_pq}), 
so we use $|\alpha~;z\!>_{\A}$ as a building block for  
the $\theta=\theta_{\psub,\qsub}$ boundary state. 
Writing it out 
explicitly,
\begin{eqnarray}
|\alpha;z>_{\A}&=&\frac1{2^{\frac52} 3^{\frac14}}
\sum_{s_0,l_j,m_j,s_j}^\beta
B^{\lambda,\mu}_\alpha
|s_0\!>\!\!>
\prod_{j=1}^3|l_j,m_j,s_j;z_j\!>\!\!>_{\A},
\label{|alpha;z>}
\end{eqnarray}
where $B^{\lambda,\mu}_\alpha$ is given by (\ref{B}).
In this case the partition function simply reduces to 
\begin{eqnarray}
&&Z^A_{(\alpha,\tilde{z})(\alpha,z)}(\tau)\nonumber\\
&&~~~~~~=
 q^{\frac16\sum_{j=1}^3 (z_j-\tilde{z}_j)^2}
\cdot\!\!\!\!\!\!\sum_{\nu_1,\nu_2,\nu_3\in \ZZsub_2}
\mbox{Orbit}\left(2-2\sum_{j=1}^3 \nu_j
;(0,0,2\nu_j);\tau(z_j-\tilde{z}_j)\right),
\label{ZAalphaz}
\end{eqnarray}
which becomes identical (up to an overall factor) to each 
spectral-flowed orbit in the summation (\ref{Z_pq}),  
provided that $z_j$ are set equal to
$z_j(\Delta m)$ (\ref{zjDeltam}).  
Fixing the normalization correctly, we find that
\begin{eqnarray}
|\theta=\theta_{\psub,\qsub}>
&\equiv&\frac1{\sqrt{2(\p^2+\p\q+\q^2)}}
\sum_{\Delta m\in \ZZsub_{\ptiny^2+\ptiny \qtiny+\qtiny^2}}
|\alpha~;z(\Delta m)>_{\A}
\end{eqnarray}
precisely yields 
%
\begin{eqnarray}
Z_{\theta=\theta_{\ptiny,\qtiny}}^{\rm open}
(\tau)
&=&
<\theta=\theta_{\psub,\qsub}|~\tilde{q}^{L_0-\frac c{24}}~
|\theta=\theta_{\psub,\qsub}>.
\end{eqnarray}
and therefore represents a D-brane tilted at an angle of
$\theta_{\psub,\qsub}$. 
Thus we have obtained, starting 
from the Recknagel-Schomerus states, the new 
ones reproducing 
the geometric D1-brane partition functions at arbitrary angles.

\subsection{Projected RS boundary states}
~~~
Let us examine the relation of our new boundary states to the 
Recknagel-Schomerus states  more closely. 
As we saw in subsect.{\ref{Istates}}, each state 
contained in  \mbox{$|l,m,s~;z>\!\!>_{\A}$} acquires a phase factor of
$e^{2\pi i zQ}$, depending on its $N=2$ $U(1)$ charge $Q$. 
If we define 
\begin{eqnarray}
P_{\psub,\qsub}|\lambda,\mu>\!\!>_{\A}
&\equiv&\frac1{\p^2+\p\q+\q^2}
\sum_{\Delta m\in \ZZsub_{\ptiny^2+\ptiny \qtiny+\qtiny^2}}
|\lambda,\mu~;z(\Delta m)>\!\!>_{\A}
\nonumber\\
&=&\frac1{\p^2+\p\q+\q^2}
\sum_{\Delta m\in \ZZsub_{\ptiny^2+\ptiny \qtiny+\qtiny^2}}
|s_0\!>\!\!>
\prod_{j=1}^3|l_j,m_j,s_j~;z_j(\Delta m)\!>\!\!>_{\A} 
\nonumber\\
\label{P_pq}
\end{eqnarray}
for $|\lambda,\mu>\!\!>_{\A}$ satisfying the $\beta$-condition, 
then $P_{\psub,\qsub}$ acts as a projection operator 
selecting only states such that 
\begin{eqnarray}
\sum_{j=1}^3Q_j\cdot z_j(\Delta m=1)\in \ZZ,
\label{projection_condition}
\end{eqnarray}
where 
\begin{eqnarray}
Q_j&=&\frac{m_j}3-\frac{s_j}2+2n_j~~~~~~(n_j\in \ZZ)
\end{eqnarray}
is the $U(1)$ charge of the $j$th $N=2$ minimal model.
Using this $P_{\psub,\qsub}$~, the boundary state 
$|\theta=\theta_{\psub,\qsub}>$ can be compactly written 
in the form
\begin{eqnarray}
|\theta=\theta_{\psub,\qsub}>&=&
\sqrt{\p^2+\p\q+\q^2}~P_{\psub,\qsub}|\theta=0>.
\end{eqnarray}

The $U(1)$-charge lattice points of 
$|\lambda,\mu>\!\!>_{\A}$ get `thinned out' after the projection 
(Figure 2). The projected boundary state 
$P_{\psub,\qsub}|\theta=0>$ is no longer a sum of products of 
Ishibashi states in general, unless 
\begin{eqnarray}
\frac{2\left[
(-\p-2\q)n_1+(2\p+\q)n_2+(-\p+\q)n_3
\right]}
{\p^2+\p\q+\q^2}
\in \ZZ \label{specialpq}
\end{eqnarray}
holds for any integers $n_j$.
(\ref{specialpq}) has a solution if and only if $\p^2+\p\q+\q^2=1,3$, 
which respectively correspond to
\mbox{$\theta=0,\frac\pi6$} (mod $\frac\pi3$). In these
special cases the boundary state $|\theta=\theta_{\psub,\qsub}>$
reduces to a  single ($\theta=0$) or a sum ($\theta=\frac\pi6$)
of Recknagel-Schomerus state(s), but otherwise it cannot be written 
in such forms.

The density of the $U(1)$-charge 
lattice gets thinner as the length of the brane becomes longer. 
On the contrary, the open-channel spectrum gets richer due to 
the modular transformation, which is in agreement with the physical 
interpretation.

\subsection{Proper free-field realizations}
~~~
We have shown how the geometric D-brane partition functions on a torus
are reproduced from the boundary states in the $(k=1)^3$ Gepner model.
But this is not the end of the story. Suppose that we consider the
partition function between the boundary states with {\it different} 
angles $|\theta=\theta_{\psub,\qsub}>$ and 
$|\theta=\theta_{\tilde{\psub},\tilde{\qsub}}>$. 
Although one would expect 
\begin{eqnarray}
<\theta=\theta_{\tilde{\psub},\tilde{\qsub}}
|~\tilde{q}^{L_0-\frac
c{24}}~ |\theta=\theta_{\psub,\qsub}> 
\label{<0|theta_pq>}
\end{eqnarray}
to coincide with the geometric answer, it does not yield 
the correct result 
$Z^{\mbox{\scriptsize open}}
_{\theta_{\tilde{\ptiny},\tilde{\qtiny}},\,
\theta_{\ptiny,\qtiny}}(\tau)$
(\ref{eq:openpfanglecft}) if the calculation is done 
using the formula (\ref{<lambdamu;z>}).

To gain insight into the problem, 
let us examine how the Gepner model reconstructs the 
momentum-winding  
lattices of the geometric boundary states. We consider the 
$\theta=0$ case first. Its momentum lattice is the most dense, 
and the corresponding boundary state on the Gepner-model side 
is the one given by the Recknagel-Schomerus construction, a
collection of all possible combinations of $N=2$ boundary states 
that satisfy the $\beta$-condition. Each of these $N=2$ 
A-boundary states further decomposes into a sum of free-boson 
Dirichlet states, whose momenta $\gamma_j$ $(j=1,2,3)$ lie on the 
cubic lattice
\begin{eqnarray}
(\gamma_1,\gamma_2,\gamma_3)
&=&
\frac1{\sqrt{12}} 
(\widehat{m}_1+12n_1,~\widehat{m}_2+12n_2,~\widehat{m}_3+12n_3),
\label{N=2cubic}\\
&&n_j\in
\ZZ~~~(j=1,2,3).\nonumber
\end{eqnarray}
The whole momentum lattice of the $\theta=0$ boundary state is 
the direct sum of (\ref{N=2cubic}) over 
all possible integers 
$\widehat{m}_j$ $(j=1,2,3)$ that fulfill the $\beta$-condition
\begin{eqnarray}
\sum_{j=1}^3 \widehat{m}_j&=& -3 s_0 +6~~~~~~(\mbox{mod $12$}).
\label{beta}
\end{eqnarray}
$\widehat{m}_j$ must be all even or all odd.
If a particular set of $\widehat{m}_j$ satisfy these conditions, then  
\begin{eqnarray}
(\widehat{m}_1,\widehat{m}_2,\widehat{m}_3)
+k_1(2,0,-2)+k_2(0,2,-2)
\label{generator}
\end{eqnarray}
also do for any $k_1,k_2\in \ZZ_6$.  
%
%
Thus, the two-dimensional sub-lattice consisting of 
the points on the plane
\begin{eqnarray}
\sum_{j=1}^3\gamma_j&=&0
\end{eqnarray}
is given by 
\begin{eqnarray}
(\gamma_1,\gamma_2,\gamma_3)
=
\frac1{\sqrt{12}} 
\left((\widehat{m}^{(0)}_1,\widehat{m}^{(0)}_2,\widehat{m}^{(0)}_3)
+k_1(2,0,-2)+k_2(0,2,-2)
\right), 
\label{2dlattice}
\end{eqnarray}
where $k_1,k_2$ may now take any integer values. 
$(\widehat{m}^{(0)}_1,\widehat{m}^{(0)}_2,\widehat{m}^{(0)}_3)$
is some reference lattice point, $(0,0,0)$ for instance.
It is then easy to check that the shift 
\begin{eqnarray}
(\widehat{m}^{(0)}_1,\widehat{m}^{(0)}_2,\widehat{m}^{(0)}_3)
\rightarrow
(\widehat{m}^{(0)}_1,\widehat{m}^{(0)}_2,\widehat{m}^{(0)}_3)
+\nu ~(1,~1,~1)
~~~~~~(\nu\in \ZZ)
\end{eqnarray}
exhausts all the points of the $\theta=0$ momentum lattice. 
The $\theta=0$ lattice is, thus, a direct product of the 
one-dimensional lattice 
\begin{eqnarray}
(\gamma_1,\gamma_2,\gamma_3)
=
\frac{\nu}{\sqrt{12}} 
(1,~1,~1),~~~~~~(\nu\in \ZZ)
\label{totalU(1)lattice}
\end{eqnarray}
and the two-dimensional one (\ref{2dlattice}),
which are orthogonal to each other.

Of course, this decomposition of the momentum lattice just 
rephrases the multiplication formula for the theta functions 
in section 4, though we are now working in the closed channel.
The extracted one-dimensional lattice 
(\ref{totalU(1)lattice}) is nothing but the $N=2$ total 
$U(1)$-charge lattice, yielding the free-fermion theta 
in the partition function. Also, the two-dimensional 
lattice (\ref{2dlattice}) coincides with the 
momentum-winding lattice of the compact bosons on the torus.
The latter can be verified as follows:
Let $\phi^j_{\theta=0}(z)$ $(j=1,2,3)$ be the free bosons that 
realize the $k=1$ minimal models and describe the boundary state
for the $\theta=0$ D-brane in the Gepner model.  
Change the orthogonal basis of the
free bosons as
\begin{eqnarray}
\left(\begin{array}{c}\varphi^1_{\theta=0}\\
\varphi^2_{\theta=0}\\ 
\varphi^3_{\theta=0}
\end{array}\right)&\equiv&
\left(\begin{array}{rcr}
\frac1{\sqrt{6}}&-\frac2{\sqrt{6}}&\frac1{\sqrt{6}}\\
\frac1{\sqrt{2}}&0&-\frac1{\sqrt{2}}\\
\frac1{\sqrt{3}}&\frac1{\sqrt{3}}&\frac1{\sqrt{3}}
\end{array}
\right)
\left(\begin{array}{c}\phi^1_{\theta=0}
\\ \phi^2_{\theta=0}
\\ \phi^3_{\theta=0}
\end{array}
\right).\label{changeofbasis}
\end{eqnarray}
$\varphi^3_{\theta=0}$ is the total 
$U(1)$ current. We also define 
$\overline\phi^j_{\theta=0}(z)$ and $\overline\varphi^j_{\theta=0}(z)$
$(j=1,2,3)$ as their anti-holomorphic counterpart, being related by 
the same transformation as (\ref{changeofbasis}).
The vertex operator of the bosons
$\phi^j_{\theta=0}(z)$ carrying the momenta (\ref{2dlattice}) 
(with $(\widehat{m}^{(0)}_1,\widehat{m}^{(0)}_2,\widehat{m}^{(0)}_3)$
\hspace{0.2cm}\hspace{-0.2cm}$=$\hspace{0.2cm}\hspace{-0.2cm}
$(\nu,\nu,\nu)$)
becomes
\begin{eqnarray}
\exp\left(i\sum_{j=1}^3\gamma_j\phi^j_{\theta=0}\right)
=\exp\left(i\sqrt{\frac23}
\left(
-\frac{\sqrt{3}}2 k_2\varphi^1_{\theta=0}
+\left(k_1+\frac{k_2}2\right)\varphi^2_{\theta=0}
\right)+i\frac{\nu}2\varphi^3_{\theta=0}
\right)
\end{eqnarray}
($k_1,k_2\in\ZZ$), and similarly for the 
anti-holomorphic bosons.
Thus, $\varphi^3_{\theta=0}$ and $\overline\varphi^3_{\theta=0}$
indeed are the bosonized complex fermions. 
On the other hand, the momentum-winding lattice of the 
compact bosons 
\begin{eqnarray}
(\widehat{X}^1_{\theta=0,L},\widehat{X}^2_{\theta=0,L};
\widehat{X}^1_{\theta=0,R},\widehat{X}^2_{\theta=0,R})
\end{eqnarray}
is 
\begin{eqnarray}
\sqrt{\frac23}
\left(\frac{\sqrt{3}}2 m,n+\frac m2;-\frac{\sqrt{3}}2 m,n
     +\frac m2
\right)  \quad (m,n \in \ZZ)      
\end{eqnarray}
($\Lambda_{1,0}$ (\ref{eq:momwindlat})), 
so this allows us to identify 
\begin{eqnarray}
\left\{
\begin{array}{rcr}
\widehat{X}^1_{\theta=0,L}&=&-~\varphi^1_{\theta=0},\\
\widehat{X}^2_{\theta=0,L}&=&\varphi^2_{\theta=0},
\end{array}
\right.~~~
\left\{
\begin{array}{rcl}
\widehat{X}^1_{\theta=0,R}&=&\overline\varphi^1_{\theta=0},\\
\widehat{X}^2_{\theta=0,R}&=&\overline\varphi^2_{\theta=0}.
\end{array}
\right.
\label{T-dual}
\end{eqnarray}
Thus we see that
the three free bosons of the $(k=1)^3$ Gepner model combine 
into a single complex free fermion and a pair of T-dual 
coordinates of the torus. The minus sign in (\ref{T-dual}) 
arises because the Neumann-Dirichlet boundary states are 
described here by only the A-boundary states.

We next consider the lattice for a general angle 
$\theta_{\psub,\qsub}$. As we saw in the previous 
subsection, it is obtained from the $\theta=0$ lattice by 
the projection (\ref{P_pq}). Leaving only the points that 
satisfy (\ref{projection_condition}) yields the lattice 
\begin{eqnarray}
(\gamma_1,\gamma_2,\gamma_3)
=
\frac1{\sqrt{12}} 
\left((\widehat{m}^{(0)}_1,\widehat{m}^{(0)}_2,\widehat{m}^{(0)}_3)
+2l_1(\p,\q,-\p-\q)+2l_2(-\q,\p+\q,-\p)
\right)
\label{2dlattice_projected}
\end{eqnarray}
($l_1,l_2\in\ZZ$). 
The vertex operator reads
\begin{eqnarray}
& &\hspace{-0.8cm}\exp\left(i\sum_{j=1}^3\gamma_j
\phi^j_{\theta=\theta_{\ptiny,\qtiny}}\right)
\nonumber\\
& &\hspace{-0.8cm}=
\exp\left(i\sqrt{\frac{2(\p^2+\p\q+\q^2)}3}
\left(
-\frac{\sqrt{3}}2 l_2\varphi^1_{\theta=\theta_{\ptiny,\qtiny}}
+\left(l_1+\frac{l_2}2\right)
                     \varphi^2_{\theta=\theta_{\ptiny,\qtiny}}
\right)+i\frac{\nu}2
\varphi^3_{\theta=\theta_{\ptiny,\qtiny}}
\rule{0mm}{7mm}\right),~~~ 
\label{vertex}
\end{eqnarray}
where $\phi^j_{\theta=\theta_{\ptiny,\qtiny}}$ $(j=1,2,3)$ are again 
the free bosons in which the $N=2$ minimal models are realized, 
but here the distinction from those for the $\theta=0$ D-brane 
is already anticipated.  $\varphi^j_{\theta=\theta_{\ptiny,\qtiny}}$ are
given by the relation
\begin{eqnarray}
\left(\begin{array}{c}\varphi^1
_{\theta=\theta_{\ptiny,\qtiny}}\\
\varphi^2_{\theta=\theta_{\ptiny,\qtiny}}\\ 
\varphi^3_{\theta=\theta_{\ptiny,\qtiny}}
\end{array}\right)\equiv
\left(
\begin{array}{rrc}
\cos\theta_{\psub,\qsub}&-\sin\theta_{\psub,\qsub}
&\\ \sin\theta_{\psub,\qsub}&\cos\theta_{\psub,\qsub}&\\
&&1
\end{array}
\right)
\left(\begin{array}{rcr}
\frac1{\sqrt{6}}&-\frac2{\sqrt{6}}&\frac1{\sqrt{6}}\\
\frac1{\sqrt{2}}&0&-\frac1{\sqrt{2}}\\
\frac1{\sqrt{3}}&\frac1{\sqrt{3}}&\frac1{\sqrt{3}}
\end{array}
\right)
\left(\begin{array}{c}\phi^1_{\theta=\theta_{\ptiny,\qtiny}}
\\ \phi^2_{\theta=\theta_{\ptiny,\qtiny}}
\\ \phi^3_{\theta=\theta_{\ptiny,\qtiny}}
\end{array}
\right).\nonumber
\\
\label{changeofbasis_pq}
\end{eqnarray} 
The anti-holomorphic bosons 
$\overline\phi^j_{\theta=\theta_{\ptiny,\qtiny}}$,
$\overline\varphi^j_{\theta=\theta_{\ptiny,\qtiny}}$  satisfy the same
relation. To compare with the geometric momentum-winding lattice at 
$\theta=\theta_{\psub,\qsub}$, we go to the local Lorentz
frame (\ref{eq:oscrotation}). Writing the lattice $\Lambda_{\psub,\qsub}$ 
(\ref{eq:momwindlat}) as the 
momentum-winding lattice of the bosons 
\begin{eqnarray}
(\widehat{X}^1_{\theta=\theta_{\ptiny,\qtiny},L},
\widehat{X}^2_{\theta=\theta_{\ptiny,\qtiny},L};
\widehat{X}^1_{\theta=\theta_{\ptiny,\qtiny},R},
\widehat{X}^2_{\theta=\theta_{\ptiny,\qtiny},R})
\end{eqnarray}
in this frame, we obtain
\begin{eqnarray}
\sqrt{\frac{2(\p^2+\p\q+\q^2)}3}
\left(\frac{\sqrt{3}}2 m,n+\frac m2;-\frac{\sqrt{3}}2 m,n+\frac m2
\right)
\end{eqnarray}
($m,n\in\ZZ$).
Therefore, comparing with (\ref{vertex}),  
we may again identify the geometric and the
Gepner-model  bosons through the relations
\begin{eqnarray}
\left\{
\begin{array}{rcr}
\widehat{X}^1_{\theta=\theta_{\ptiny,\qtiny},L}
&=&-~\varphi^1_{\theta=\theta_{\ptiny,\qtiny}}~,\\
\widehat{X}^2_{\theta=\theta_{\ptiny,\qtiny},L}
&=&\varphi^2_{\theta=\theta_{\ptiny,\qtiny}}~,
\end{array}
\right.~~~
\left\{
\begin{array}{rcl}
\widehat{X}^1_{\theta=\theta_{\ptiny,\qtiny},R}
&=&\overline\varphi^1_{\theta=\theta_{\ptiny,\qtiny}}~,\\
\widehat{X}^2_{\theta=\theta_{\ptiny,\qtiny},R}
&=&\overline\varphi^2_{\theta=\theta_{\ptiny,\qtiny}}~.
\end{array}
\right.
\label{T-dual_pq}
\end{eqnarray}

We are at last in a position to understand why the naive 
evaluation of (\ref{<0|theta_pq>}) did not give 
$Z^{\mbox{\scriptsize
open}}_{\theta_{\tilde{\ptiny},\tilde{\qtiny}},\,
\theta_{\ptiny,\qtiny}}(\tau)$
(\ref{eq:openpfanglecft}). Owing to the sign-flip which occurs in the 
left-moving part (\ref{T-dual_pq}), the free bosons 
$\varphi^a_{\theta=\theta_{\ptiny,\qtiny}}$ and 
$\overline{\varphi}^a_{\theta=\theta_{\ptiny,\qtiny}}$ 
$(a=1,2)$ rotate in the {\it opposite} directions
\begin{eqnarray}
\varphi^a_{\theta=\theta_{\tilde{\ptiny}
          ,\tilde{\qtiny}}}
=R(-\Delta\theta)^a_{~~b}\varphi^b_{\theta=\theta_{\ptiny,\qtiny}}~,
~~~~~~
\overline\varphi^a_{\theta=\theta_{\tilde{\ptiny},\tilde{\qtiny}}}
=R(\Delta\theta)^a_{~~b}\overline\varphi^b_{\theta=\theta_{\ptiny,\qtiny}}
\label{varphirotation}
\end{eqnarray}
with
\begin{eqnarray}
R(\Delta\theta)^a_{~~b}=\left(
\begin{array}{rr}
\cos\Delta\theta &\sin\Delta\theta\\
-\sin\Delta\theta&\cos\Delta\theta
\end{array}
\right),~~~\Delta\theta=\theta_{\tilde{\psub},\tilde{\qsub}}
-\theta_{\psub,\qsub}~.
\end{eqnarray}
Therefore, if $\theta_{\psub,\qsub}
\neq\theta_{\tilde{\psub},\tilde{\qsub}}$~, 
an A-boundary state written in 
$\varphi^a_{\theta=\theta_{\ptiny,\qtiny}}$, 
$\overline\varphi^a_{\theta=\theta_{\ptiny,\qtiny}}$ does not have 
equal left-right momenta in 
$\varphi^a_{\theta=\theta_{\tilde{\ptiny},\tilde{\qtiny}}}$, 
$\overline\varphi^a_{\theta=\theta_{\tilde{\ptiny},\tilde{\qtiny}}}$, 
except for the zero-momentum state. 
Another striking consequence of the relations 
(\ref{varphirotation}) is that
the free  bosons $\phi^j_{\theta=\theta_{\ptiny,\qtiny}}$, 
$\overline\phi^j_{\theta=\theta_{\ptiny,\qtiny}}$ realizing  
individual $N=2$ minimal models are {\it not the same} if  the angle 
of the D-brane is different.  
To distinguish one tilted D-brane from
another,  one must specify which free fields are used to realize 
the $N=2$ minimal models!

Taking into account the rotation (\ref{varphirotation}), we now 
write
\begin{eqnarray}
|\theta=\theta_{\psub,\qsub}>&=&
\sqrt{\p^2+\p\q+\q^2}~P_{\psub,\qsub}|\theta=0>^{\!
\theta_{\ptiny,\qtiny}},
\end{eqnarray}
where by $|\theta=0>^{\!
\theta_{\ptiny,\qtiny}}$ we denote the previous $\theta=0$ 
boundary state but made up of the particular realization
$\phi^j_{\theta=\theta_{\ptiny,\qtiny}}$, 
$\overline\phi^j_{\theta=\theta_{\ptiny,\qtiny}}$ 
($j=1,2,3$). We also write the product of Ishibashi states 
$|\lambda,\mu~;z>\!\!>^{\theta_{\ptiny,\qtiny}}$  
in the same meaning.
Then the amplitude 
\begin{eqnarray}
{}^{\theta_{\tilde{\ptiny},\tilde{\qtiny}}}
\!\!<\!\!<\lambda,\mu~;\tilde{z}~|
~\tilde{q}^{L_0-\frac c{24}}~
|\lambda,\mu~;z>\!\!>^{\theta_{\ptiny,\qtiny}}
\label{<lambdamu_tilted>}
\end{eqnarray}
is not just a product of characters, but instead given, 
for $(\lambda, \mu)=(s_0;(l_j,m_j,s_j))$,
by
\begin{eqnarray}
(\ref{<lambdamu_tilted>})
=\chi^{SO(6)}_{s_0}(\tilde{\tau})\cdot
\frac{\tilde{q}^{-\frac18}}{\prod_{n=1}^\infty\left((1-\tilde{q}^n
e^{2i\Delta\theta}) (1-\tilde{q}^ne^{-2i\Delta\theta})
\right)}\cdot
\frac{\Theta_{-3s_0+6+12r,~18}(\tilde\tau,\frac{\Delta\theta}{3\pi})}
{\prod_{n=1}^\infty(1-\tilde{q}^n)},
\label{result<lambdamu_tilted>}
\end{eqnarray}
where 
\begin{eqnarray}
\sum_{j=1}^3\widehat{m}_j=-3s_0+6+12r,
~~~\widehat{m}_j=2m_j-3s_j,
~~~r\in \ZZ_3.
\end{eqnarray}
The second factor of (\ref{result<lambdamu_tilted>}) comes from the
oscillator excitations of  
$(\varphi^a_{\theta=\theta_{\ptiny,\qtiny}},
\overline\varphi^a_{\theta=\theta_{\ptiny,\qtiny}})$ and 
$(\varphi^a_{\theta=\theta_{\tilde{\ptiny},\tilde{\qtiny}}},
\overline\varphi^a_{\theta=\theta_{\tilde{\ptiny},\tilde{\qtiny}}})$ 
$(a=1,2)$. 
By construction,
the calculations are the same as before
for the compact bosons.
Also, we have no theta here because the 
momentum overlap occurs only at one point. The $z$ or
$\tilde{z}$ dependence thus disappears.
The last factor is the contribution from 
$(\varphi^3_{\theta=\theta_{\ptiny,\qtiny}},
\overline\varphi^3_{\theta=\theta_{\ptiny,\qtiny}})$ and 
$(\varphi^3_{\theta=\theta_{\tilde{\ptiny},\tilde{\qtiny}}},
\overline\varphi^3_{\theta=\theta_{\tilde{\ptiny},\tilde{\qtiny}}})$.  
The denominator is from the ordinary oscillator contribution, 
whereas the theta function comes from the momentum overlap of 
$\varphi^3_{\theta=\theta_{\ptiny,\qtiny}}$ and 
$\varphi^3_{\theta=\theta_{\tilde{\ptiny},\tilde{\qtiny}}}$. 
Here we have identified these bosons through the 
modulus shift 
\begin{eqnarray}
\varphi^3_{\theta=\theta_{\tilde{\ptiny},\tilde{\qtiny}}}
+\overline\varphi^3_{\theta=\theta_{\tilde{\ptiny},\tilde{\qtiny}}}
&=&\varphi^3_{\theta=\theta_{\ptiny,\qtiny}}
+\overline\varphi^3_{\theta=\theta_{\ptiny,\qtiny}}
+2\Delta\theta 
\label{varphi3shift}
\end{eqnarray}
in order to take into account the rotation of the compact 
fermions, so that the level-18 theta acquires 
$\Delta\theta/3\pi$. Thus, in all, we obtain the equality 
(\ref{result<lambdamu_tilted>}). Using this formula, we calculate 
the partition function as
\begin{eqnarray}
&&<\theta=\theta_{\tilde{\psub},\tilde{\qsub}}
|~\tilde{q}^{L_0-\frac
c{24}}~ |\theta=\theta_{\psub,\qsub}> \nonumber\\
&&~~~~=\sqrt{\tilde{\p}^2+\tilde{\p}\tilde{\q}
+\tilde{\q}^2}\sqrt{\p^2+\p\q+\q^2} \,\,\,
{}^{\theta_{\tilde{\ptiny},\tilde{\qtiny}}}\!<\theta=0
|~\tilde{q}^{L_0-\frac
c{24}}~ |\theta=0>^{\theta_{\ptiny,\qtiny}}\nonumber\\
&&~~~~=\sqrt{\tilde{\p}^2+\tilde{\p}\tilde{\q}
+\tilde{\q}^2}\sqrt{\p^2+\p\q+\q^2}
\left(
\frac{1}{\sqrt{2}}\right)^2
\left(
\frac{2^3}{2^{\frac52}3^{\frac14}}\right)^2
\nonumber\\
&&~~~~~~~~~\cdot\sum_{s_0\in \ZZsub_4}\sum_{r\in \ZZsub_3}
(-1)^{s_0^2}
\cdot
\frac{\tilde{q}^{-\frac18}
\chi^{SO(6)}_{s_0}(\tilde{\tau})
\Theta_{-3s_0+6+12r,~18}(\tilde\tau,\frac{\Delta\theta}{3\pi})
}{\prod_{n=1}^\infty\left((1-\tilde{q}^n
e^{2i\Delta\theta}) (1-\tilde{q}^ne^{-2i\Delta\theta})
(1-\tilde{q}^n)\right)}\nonumber \\
&&~~~~=\sqrt{\frac{(\tilde{\p}^2+\tilde{\p}\tilde{\q}
+\tilde{\q}^2)(\p^2+\p\q+\q^2)}3}\nonumber \\
&&~~~~~~~~~\cdot\sum_{s_0\in \ZZsub_4}
(-1)^{s_0^2}\cdot
2\sin\Delta\theta\cdot \frac{\chi^{SO(6)}_{s_0}(\tilde{\tau})
\Theta_{-s_0+2,2}(\tilde\tau,\frac{\Delta\theta}{\pi})}
{\vartheta_1(\tilde\tau,\frac{\Delta\theta}{\pi})}.
\label{finalexpression}
\end{eqnarray}
The projection operator $P_{\psub,\qsub}$  
drops out since 
it is inert for the states having zero 
$\varphi^a_{\theta=\theta_{\ptiny,\qtiny}}$ 
momenta $(a=1,2)$. By a modular transformation, the final 
expression (\ref{finalexpression}) precisely reproduces 
$Z^{\mbox{\scriptsize open}}
_{\theta_{\tilde{\ptiny},\tilde{\qtiny}},\,
\theta_{\ptiny,\qtiny}}(\tau)$
(\ref{eq:openpfanglecft}) in section 3.

\subsection{The Witten index}
~~~
The angle-dependent identification (\ref{varphirotation}) is  
also essential for the Witten index to give the correct intersection
numbers of the  cycles. The open-string Witten index for the D-branes
tilted  at angles of $\theta=\theta_{\psub,\qsub}$ and
$\theta_{\tilde{\psub},\tilde{\qsub}}$ 
($(\p,\q)\neq (\tilde{\p},\tilde{\q})$) is defined by the
transition amplitude  between the corresponding boundary states with the 
$(-1)^{F_L}$ insertion in the closed channel
\begin{eqnarray}
I_{\theta_{\tilde{\ptiny},\tilde{\qtiny}},\,\theta_{\ptiny,\qtiny}}
&\equiv&{}_{\mbox{\scriptsize R,int}\!}
<\theta=\theta_{\tilde{\psub},\tilde{\qsub}}|
(-1)^{F_L}
\tilde{q}^{L_0-\frac c{24}}
|\theta=\theta_{\psub,\qsub}>_{\mbox{\scriptsize R,int}}
,
\end{eqnarray}
where R,int indicates that we only consider the internal part 
of the boundary state in the Ramond sector. 
The total fermion number is
\begin{eqnarray}
F_L&=&\frac12+\frac16\sum_{j=1}^3\widehat{m}_j~~~\in \ZZ.
\end{eqnarray}
Using (\ref{result<lambdamu_tilted>}) 
with the $SO(6)$ character being removed, 
$I_{\theta_{\tilde{\ptiny},\tilde{\qtiny}},\,
\theta_{\ptiny,\qtiny}}$
is similarly calculated as
\begin{eqnarray}
&I_{\theta_{\tilde{\ptiny},\tilde{\qtiny}},\,
\theta_{\ptiny,\qtiny}}&
=\sqrt{\tilde{\p}^2+\tilde{\p}\tilde{\q}
+\tilde{\q}^2}\sqrt{\p^2+\p\q+\q^2}~\,
{}^{\theta_{\tilde{\ptiny},\tilde{\qtiny}}}
_{\mbox{\scriptsize R,int}}
\hspace{-0.1cm}<\theta=0
|(-1)^{F_L}\tilde{q}^{L_0-\frac
c{24}}~ |\theta=0>^{\theta_{\ptiny,\qtiny}}
_{\mbox{\scriptsize R,int}}
\nonumber\\
&&=\sqrt{\tilde{\p}^2+\tilde{\p}\tilde{\q}
+\tilde{\q}^2}\sqrt{\p^2+\p\q+\q^2}
\left(
\frac{1}{\sqrt{2}}\right)^2
\left(
\frac{2^3}{2^{\frac52}3^{\frac14}}\right)^2
\nonumber\\
&&\cdot\sum_{s_0=\pm 1}\sum_{r\in \ZZsub_3}
(-1)^{s_0^2+\frac12+\frac16\sum_{j=1}^3\widehat{m}_j}
\cdot
\frac{\tilde{q}^{-\frac18}
\Theta_{-3s_0+6+12r,~18}(\tilde\tau,\frac{\Delta\theta}{3\pi})
}{\prod_{n=1}^\infty\left((1-\tilde{q}^n
e^{2i\Delta\theta}) (1-\tilde{q}^ne^{-2i\Delta\theta})
(1-\tilde{q}^n)\right)}\nonumber \\
&&=\sqrt{\frac{(\tilde{\p}^2+\tilde{\p}\tilde{\q}
+\tilde{\q}^2)(\p^2+\p\q+\q^2)}3}
\cdot 2i\sin\Delta\theta
\nonumber \\
&&=i(\tilde{\q}~\p-\tilde{\p}\q),
\end{eqnarray}
which is precisely ($i$ times) the intersection number of two wound 
branes on the torus. If we take the total (internal) $U(1)$ 
charge $J_0^{N=2}=\frac16\sum_{j=1}^3\widehat{m}_j$ in place of 
$F_L$ in the definition, we get an integer index without $i$.

\setcounter{equation}{0}
\section{Comments on the $(k=2)^2$ model}
~~~
In this section, we briefly outline how to construct 
in the $(k=2)^2$ Gepner model the boundary states
for arbitrary $(\p,\q)$ branes on the corresponding
$SU(2)^2$ torus.
In this case, 
the geometric partition function for parallel branes
(\ref{eq:openpfpqcft}) is given by 
\beqa
Z^{\mbox{\scriptsize{open}}}_{\theta=\theta_{\ptiny,\qtiny}}
(\tau)
&=&(\mbox{Jacobi})\cdot\frac{1}{\eta^2 (\tau)}
   \sum_{m,n \in \ZZsub}
   q^{\frac{m^2+n^2}{\psub^2+\qsub^2}}     \non \\
&=&\sum_{\Delta m \in \ZZsub_{\ptiny^2+\qtiny^2}}
      q^{\frac{(\Delta m)^2}{\psub^2+\qsub^2}}
  \sum_{\nu \in \ZZsub_4}(-1)^{\nu}\chi_{-\nu}^{SO(6)}(\tau)
                                                      \non \\  
& & \hspace{-1.0cm} \cdot 
 \frac{1}{\eta^3 (\tau)}
 \Theta_{\nu+2,2}(\tau,0)
 \sum_{s \in \ZZsub_2}\Theta_{2s,2}
 (\tau, \frac{(\p-\q)\tau \Delta m}{\p^2+\q^2})
                 \Theta_{2s,2}(\tau, \frac{(\p+\q)\tau \Delta m}
                                          {\p^2+\q^2}).    
\label{eq:22pq}
\eeqa
As in the case of the $SU(3)$ torus,
this partition function can be thought of
as obtained from a suitable projection of 
the $\theta=0$ partition function in the closed-string channel.

Let us discuss the corresponding projection in the 
Gepner model.
It was found \cite{GS} that
the partition function for $\theta=0$ ($\p=1,\q=0$)
is reproduced from the amplitude (\ref{Z_alphaalpha}) for the 
Recknagel-Schomerus boundary states.
This is shown by rearranging the product 
of two $k=2$ minimal model characters 
\beq
 \chi^{l,s}_m(\tau,z)=c^l_{m-s}(\tau)
                    \Theta_{m-2s,4}(\tau,\frac{z}{2})
\eeq 
in the amplitude as a triple product of theta functions.
One of the last two theta's in (\ref{eq:22pq})
comes from the product of the string functions 
$c^l_{m-s}(\tau)$,
which are essentially the partition functions for 
the parafermions in the free-field realizations.
The important point is, 
for a general $\theta_{\psub,\qsub}$,  
that the $\Delta m$ dependence in those theta's
results in a particular projection in the closed-string channel
after summing over $\Delta m$. 
Therefore,
in the $(k=2)^2$ case, we see that
not only the $U(1)$ charges but also
the parafermions are subject to the projection.

The easiest way to handle such a projection is 
to bosonize the parafermions, which are real free
fermions in this model. 
Once we rewrite the Recknagel-Schomerus
boundary state for $\theta=0$ in terms of the free bosons,
the boundary state for $\theta_{\psub,\qsub}$
is obtained by the projection similarly
to the $(k=1)^3$ case. 
The bosonization  
also enables us to specify the free-field realizations, 
which is needed to reproduce the partition functions
at angles.
The details of the construction will be reported elsewhere.

\section{Conclusions}
~~~
We found that, to represent infinitely many wound 
D1-branes in the ($k$=1)$^3$ Gepner model, we can no longer keep, in
general, the  `structure' of $N=2$ Ishibashi states. We need to project
out some states to have a rich spectrum in the open channel, but then 
this forces us to give up imposing any fixed A-boundary conditions 
on the individual $N=2$ minimal models. The projected boundary state 
is obtained by writing the Recknagel-Schomerus states in terms of 
free-boson boundary states with  moduli and summing over 
shifts.  This shift is orthogonal to the total $U(1)$ charge, and
therefore does not break spacetime supersymmetry. Each boundary state in
the summation can be regarded as an $N=2$ A-boundary state in some
realization, but the projected state as a whole is not.

Another unexpected aspect of the new boundary states is the necessity 
of the proper realization for each angle of the D-brane. 
We found it necessary, in particular, for the Witten index to correctly
yield the intersection numbers, at least for the Gepner-model descriptions
of the toroidal compactifications. After all, to describe infinitely 
many supersymmetric D-branes, we need more information than just an
algebraic $N=2$ Ishibashi state has, that is, in which free fields it is
realized.

One might then ask, ``Why have the consistent intersection numbers been
obtained so far without taking into account such `angle-wise' realizations
in the literature?''  Our explanation for this is that  the parafermions
in those models carry some information on the angles of the branes, and
probably the previous discussions were  the special cases in which no
distinction among the realizations  was necessary.

It would be extremely interesting to study whether a similar
projection in general Gepner models yields new boundary states 
representing D-branes wound around more general supersymmetric cycles of
the Calabi-Yau. We again start from a Recknagel-Schomerus boundary state.
The guiding principle is that its constituent  states get projected and
thinned out depending on the $N=2$
$U(1)$  charges they carry. The resulting boundary state is supersymmetric 
if and only if the shift of the $U(1)$-boson modulus is orthogonal 
to the total $U(1)$ current. To proceed further, we must consider 
the following:

First, we need to have some convenient realization for the parafermions 
to systematically project out some of the states from the parafermion
state space. The parafermion piece of the $N=2$ free fields becomes
trivial if $k=1$, which allowed us in the $(k=1)^3$ model to only
consider the projection for the $U(1)$ bosons. As is already clear in the
$(k=2)^2$  case, not only the free-boson piece but also the parafermion
piece of the states are subject to the projection, in a correlated
manner. In the $(k=2)^2$ example, the two real fermions (= $\ZZ_2$
parafermions) are conveniently bosonized, and get projected similarly to
the $(k=1)^3$ model. In general cases, the Wakimoto or the coset
realization might be of use for this purpose. Another thing we have to
worry about is how to determine the relations among the free-field
realizations for different projected boundary states. This will be hard
because we have no analogue of the toroidal CFT description in general.
One possible criterion is to require that the boundary states have
integral Witten indices among themselves. 

\section*{Acknowledgements}
~~~
We would like to thank M.~Naka and Y.~Satoh
for discussions. 
The work of T.~T. has been
supported by Soryushi Shogakukai.

\newpage


~~~ \vspace{4.5cm} \\
\begin{figure}[h]
\begin{center}
\resizebox{!}{7cm}{\includegraphics{pqcycle.eps}}
\end{center}
\centerline{Figure.1 The $(\p,\q)$ cycle.}
\end{figure}

\begin{figure}[t]
\begin{center}
\resizebox{!}{7cm}{\includegraphics{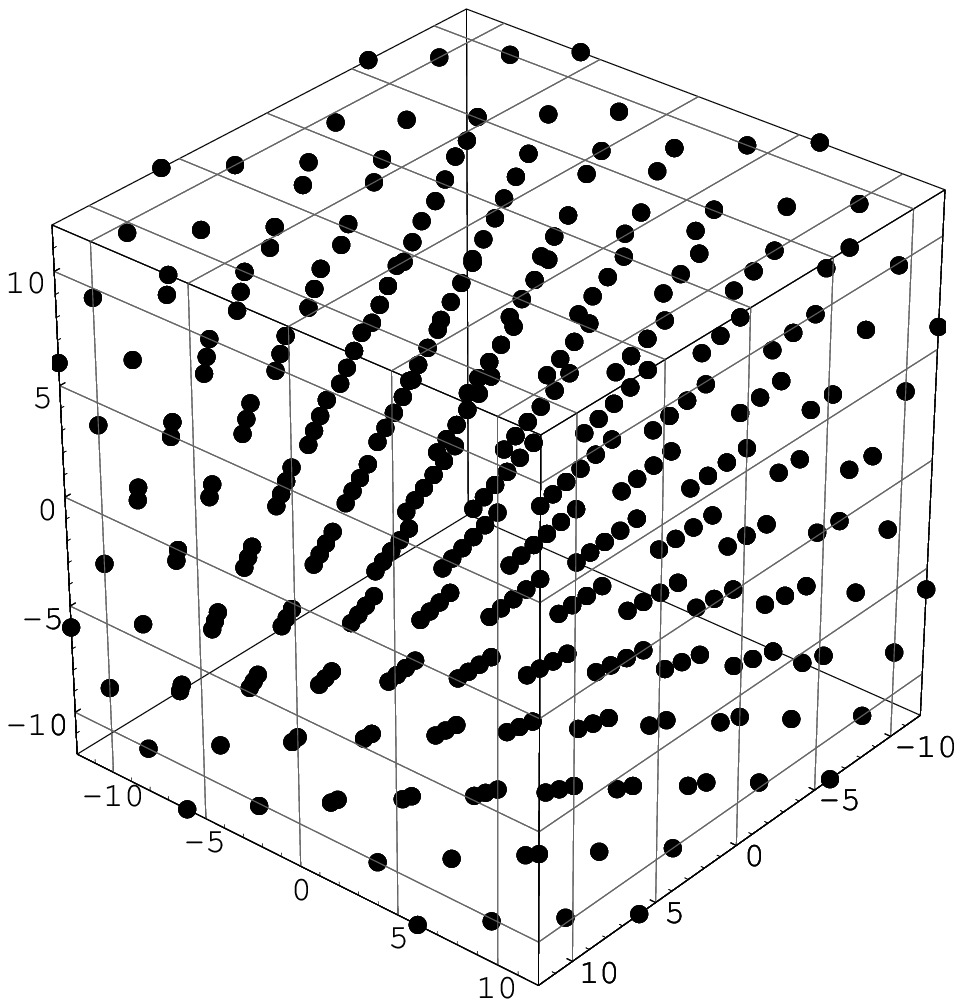}}
\end{center}
\centerline{Figure.2
The momentum lattice of the boundary state }
\centerline{for $s_0=0$ representing 
(a) the $\theta=0$ D-brane.~~~~~~~~~~~~}
\end{figure}

\begin{figure}
\begin{center}
\resizebox{!}{7cm}{\includegraphics{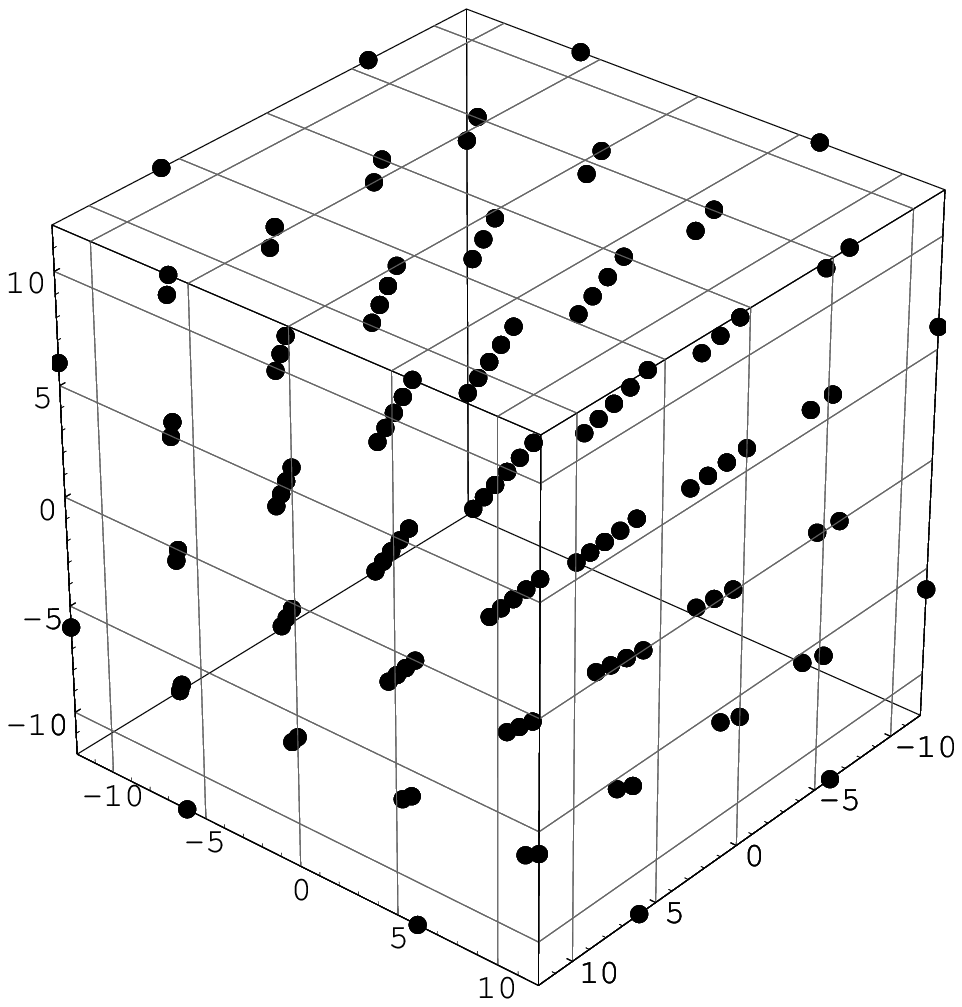}}
\end{center}
\centerline{(b) The $\theta=\frac\pi 6$ D-brane. 
The lattice points form lines parallel to }
\centerline{the total $U(1)$ direction ($\beta$-orbit). 
(The axes are scaled by 6.)}
\centerline{They are thinned out due to the projection.~~~~~~~~~~~~~~~~~~~~~~~}
\end{figure}

\end{document}